\documentclass[11pt,a4paper]{article}

\usepackage{jheppub}
\usepackage[sort&compress]{natbib}
\usepackage{url}
\usepackage{hyperref}
\usepackage{amsfonts}
\usepackage{epsfig}\usepackage{dcolumn}
\usepackage{bm}
\usepackage{slashed}
\graphicspath{{./figuras/}}

\def\br{\begin{eqnarray}}
\def\er{\end{eqnarray}}
\def\be{\begin{equation}}
\def\ee{\end{equation}}
\def\to{\rightarrow}

\newcommand{\gev}{\; \hbox{GeV}}
\newcommand{\ifb}{\; \hbox{fb}^{-1}}
\newcommand{\zp}{Z^{\prime}}

\newcommand{\mzp}{M_{Z'}}
\newcommand{\mchi}{M_{\chi}}

\newcommand{\gx}{g_{\chi}}
\newcommand{\eslash}{\not\!\!\! E_T}

\title{\mbox{The Dark $Z^{\prime}$ Portal: Direct, Indirect and Collider Searches}}

\author{Alexandre Alves$^a$,}
\author{Stefano Profumo$^b$,}
\author{Farinaldo S. Queiroz$^b$}

\affiliation{$^a$Departamento de Ci\^encias Exatas e da Terra,
Universidade Federal de S\~ao Paulo, Diadema-SP, 09972-270, Brasil}
\affiliation{$^b$Department of Physics and Santa Cruz Institute for Particle Physics
University of California, Santa Cruz, CA 95064, USA}

\emailAdd{aalves@unifesp.br}
\emailAdd{profumo@ucsc.edu}
\emailAdd{fdasilva@ucsc.edu}

\abstract{We perform a detailed study of the dark $\zp$ portal using a generic parametrization of the $\zp$-quarks couplings, both for light ($8-15$)~GeV and heavy ($130-1000$)~GeV dark matter scenarios. We present a comprehensive study of the collider phenomenology including jet clustering, hadronization, and detector artifacts, which allows us to derive accurate bounds from the search for new resonances in dijet events and from mono-jet events in the LHC $7$~TeV, LHC $8$~TeV, and Tevatron $1.96$~TeV data. We also compute the dark matter relic abundance, the relevant scattering cross sections and pair-annihilation spectrum, and compare our results with the current PLANCK, Fermi-LAT and XENON100/LUX bounds.  
Lastly, we highlight the importance of complementary searches for dark matter, and outline the excluded versus still viable parameter space regions of the dark $\zp$ portal. 
  }
    
\begin{document} 
\maketitle
\flushbottom

\section{Introduction}
\label{introduction}

The nature of the Dark Matter (DM) that permeates our Universe is one of the most mysterious puzzles in current science. A variety of dark matter candidates have been proposed and investigated. One of the most compelling dark matter candidates is the so-called WIMP (Weakly Interacting Massive Particle): WIMPs  naturally arise in many independently well-motivated theories for physics beyond the Standard Model, and produce a thermal relic density often in accordance with the observed dark matter density.

Recent signals from direct detection searches have brought considerable attention to a ``light'' dark matter scenario \cite{DD1,DD2,DD3,DD4,DD5,DD6,DD7,DD8,DD9,DD10}, because light ($\sim 8-15$~GeV) DM particles that interact coherently with nucleons may account for the excess events observed by DAMA, CoGENT, CRESST, CDMS-Si underground experiments \cite{DDexp1,DDexp2,DDexp3,DDexp4,DDexp5,DDexp6}.  On the other hand, recent XENON100 \cite{XENON100} and LUX \cite{LUX} null results have effectively ruled out the light dark matter window. Even the XENON-phobic scenarios discussed in \cite{xenonphobia1,xenonphobia2} do not appear to suffice anymore. Here, we will use LUX and XENON100 limits on the scattering cross section to bound the parameter space of the $Z^{\prime}$ portal.

Indirect searches for DM pair-annihilation in our Galaxy have also resulted in tentative evidence for light dark matter particles. In particular, the $\sim 1-3$ GeV gamma-ray excess observed at the Galactic center in the Fermi-LAT data can be plausibly explained either by a $\sim 8$~GeV DM particle annihilating into $\tau^+\tau^-$ or a $\sim 25$~GeV one annihilating mostly into $b\bar b$ \cite{ID1,ID2,ID3,ID4,ID5,ID6}. This excess has been reported by different groups \cite{ID1,ID2,ID3,ID4,ID5,ID6}. Astrophysical uncertainties, such as the gas distribution surrounding the Galactic Center and unaccounted-for astrophysical sources such as unidentified pulsars, however, blur the significance of this signal, which thus warrants additional critical scrutiny. A few analyses which use AMS-02 and Fermi-LAT data also seem to disfavor a light dark matter scenario \cite{danPRL,FarinaldoGC,Fermidwarfs}. Besides this controversial gamma-ray excess, the positron excess observed in the AMS-02 data \cite{AMS02} and before that in Fermi-LAT and Pamela data, which might also be explained by a $\sim 500$~GeV DM, seems to be more plausibly due to a local pulsar population \cite{profumo}. Furthermore, an exciting $130$~GeV gamma-ray line observed in the Fermi-LAT data and reported by different groups \cite{gammaline1,gammaline2,lineother}, might potentially be associated with the two-photon annihilation of a $130$~GeV DM particle, but might also reflect an instrumental effect. Be it as it may, those different indirect detection signals are quite interesting and promising and are expected to be settled in either direction in the near future. In this work we will use the current Fermi-LAT dwarfs bounds \cite{Fermidwarfs} to constrain the annihilation cross section of the WIMPs, but will refer to some of the mentioned tentative signals in our choice of dark matter particle masses.

Regarding collider searches, the basic signature of DM production is the presence of missing energy, because WIMP DM particles generically escape the detector. For this reason collider searches for DM production typically involve jets + $\eslash$ data. Colliders provide important complementary bounds mostly in the light dark matter window where direct detection experiments are threshold limited, \cite{ATLASDM1,ATLASDM2,ATLASDM3,ATLASDM4}. We will see that collider bounds are moderately sensitive to dark sector features such as the mass of the DM particle and couplings. However, the coupling strength of the $Z^{\prime}$ with the quarks is very important in setting the size of the production cross section and the hardness of the jets. In our analysis, we will derive constraints using the LHC 7 and 8 TeV data plus the Tevatron $1.96$ TeV Refs.\cite{tevjj,lhc7jj,lhc8jj}

Our goal in the present study is to outline the viable parameter space of the $Z^{\prime}$ portal by taking into account the complementarity of direct, indirect and collider searches \cite{ZpComplementary1,ZpComplementary2,ZpComplementary3,ZpComplementary4,ZpComplementary5}. Our work goes beyond and differs from previous studies in several ways, including the following: 

\begin{itemize}
\item We use an effective Lagrangian approach adding together spin dependent and independent couplings.
\item We include indirect detection bounds.
\item We perform a comprehensive collider study by including jet clustering and hadronization, and by simulating detector effects.

\item We outline the viable parameter space in the $Z^{\prime}$ mass $Z'-DM-DM$ plane, after plugging into the most relevant colliders (LHC 8,LHC 7, Tevatron $1.96$ TeV), direct (LUX,XENON100) and indirect (Fermi Dwarfs) limits.

\end{itemize}

\section{A Z$^{\prime}$ Portal Dark Matter Model}
New heavy neutral gauge bosons, $\zp$, appear in many gauge extensions of the Standard Model (SM). In particular, in certain models these additional gauge bosons are responsible for linking the dark and visible sector \cite{ZpDM1,ZpDM2,ZpDM3,ZpDM4,ZpDM5,ZpDM6,ZpDM7,ZpDM9,ZpDM10} producing a so-called $\zp$ ``{\em portal}''. In this work we aim to investigate this setup under three different perspectives: direct, indirect and collider searches. To do so, we use a general Lagrangian that reads, 
\begin{eqnarray}\label{lgrg1}
{\cal L} &=& -\frac{g}{2C_W} \left[\sum_{i} \bar{q_i} \gamma^{\mu} (a \cdot g_V^i - b\cdot g_A^i \gamma_{5}) q_i\right]Z'_{\mu}
+ g_{\chi} \left[ \bar{\chi} \gamma^{\mu} (1 -\gamma_{5}) \chi \right]  Z'_{\mu}\ ,
\label{eq1}
\end{eqnarray}where $q_i$'s are denoting the SM quarks, $a$ and $b$ are constant factors, which are equal to unity if one assume that the $Z^{\prime}$ couples equivalently to the SM $Z$ boson, whereas $\chi$ and $g_{\chi}$ are the dark matter particle and the $DM-DM-Z^{\prime}$ coupling, respectively. Here $g_V^i$ and $g_A^i$ are the vector and axial $Z$-quarks couplings, which read \cite{PDGZp},
\begin{eqnarray}
g_V^i & =& t_{3L}(i)-2\cdot Q_i S^2_W,\nonumber\\
g_A^i & =& t_{3L}(i).
\end{eqnarray}$Q_i$ is the charge of the quark i,$t_{3L}$ is the weak isospin of quarks, with $t_{3L}=+1/2\ (-1/2)$ for up (down) quarks.
From Eq.(\ref{eq1}) we see that:
\begin{itemize}
\item The $Z^{\prime}$ is Leptophobic, i.e. it just doesn't couple to leptons by construction. This is the most constrained scenario by collider searches concerning a new neutral gauge boson once the $\zp$ can only decay to jets and DM pairs. On the other hand, DM-nucleon interactions do not depend upon the details of the leptonic sector.
\item The constant factors {\it a} and {\it b} are equal to unity in the SM. Here, we will  explore different particle physics models by varying these constants. We will assume that $a=b$ and investigate two scenarios:

(i) $a=b=1$; 

(ii) $a=b=0.5$. 

The latter scenario corresponds to the case where the $Z^\prime$-quarks couplings are suppressed by $50\%$ in comparison with $Z$-quarks ones.
\item As described in Table~\ref{table1}, the Dark $Z^{\prime}$ portal might give rise to four different operators when $\chi$ is a Dirac Fermion \cite{ZpComplementary4,ZpComplementary5,ZpComplementary6,ZpComplementary7,ZpComplementary8,
ZpComplementary9}. If $\chi$ is a Majorana fermion, the vector current is zero. Therefore, in the latter case we are left with the $O_2$ and $O_4$ operators. In this work we will not study these operators individually. We consider them all at once. Therefore we assume that $\chi$ is a Dirac fermion with vector and axial coupling to fermions, as described in Eq.(\ref{eq1}).
\begin{table}[h!]
\begin{center}
\begin{tabular}{lllc}
& &{\color{blue}The Leptophobic Dark $Z^{\prime}$ Portal}&\\
\hline
~&~ {\bf Operator}                                                            ~&~ {\bf Structure} ~&~ {\bf Scattering Cross Section}\\
\hline
\rule{0cm }{0.7cm} $O_1$  \vspace{0.1cm}
 ~&~ $\bar q\gamma^\mu q \bar\chi\gamma_\mu\chi$                             ~&~ Spin Independent ~&~  $\frac{9 g_V^2 g_{\chi}^2 M_n^2 \mchi^2}{\pi \mzp^4 (M_n+\mchi)^2}$     \\
\hline
\rule{0cm }{0.7cm} $O_2$  \vspace{0.1cm}
 ~&~ $\bar q\gamma^\mu q\bar\chi\gamma_\mu\gamma_5\chi$                      ~&~ Spin Independent ~&~ $\sim v^2$ \\
\hline
\rule{0cm }{0.7cm} $O_3$  \vspace{0.1cm}
~&~ $\bar q\gamma^\mu\gamma_5 q \bar\chi\gamma_\mu\chi$                      ~&~ Spin Dependent ~&~ $\sim v^2$\\
\hline
\rule{0cm }{0.7cm} $O_4$  \vspace{0.1cm}
~&~ $\bar q\gamma^\mu\gamma_5 q\bar\chi\gamma_\mu\gamma_5\chi$              ~&~ Spin Dependent  ~&~ $\frac{3 g_A^2 g_{\chi}^2 (\Delta\Sigma)^{2} M_n^2\mchi^2}{\pi \mzp^4 (M_n+\mchi)^2}$ \\
\hline
\end{tabular}
\end{center}
\label{table1}
\caption{Effective operators for DM-Nucleon scattering. We have classified the Spin Independent (SI) and Spin Dependent (SD) operators. $v$ is the velocity of DM in the lab frame, $M_n$ is the nucleon mass and $\Delta\Sigma$ is defined as $\langle N|\sum_{q}\bar q \gamma_{\mu}\gamma_{5} q |N\rangle = \Delta\Sigma \bar U_{N}\gamma_{\mu}\gamma_{5} U_{N}$, with $U_{N}$ as the wave function of the nucleon \cite{ZpComplementary4}. The $g_V$ and $g_A$ couplings are determined according to Eq.(\ref{eq1}).}
\end{table}
\item $Z^{\prime}$ gauge bosons are predicted to exist in extended gauge theories, such as $U(1)_X$. In principle, the setup investigated here is supposed to include extra fermions to cancel the anomalies induced by the $U(1)_X$. In our scenario the mass of the extra fermion should be $M_{f} \leq 64\pi^2/(a*g_V^3) M_{Z^{\prime}}$ \cite{extraf}. However, this bound may be circumvented if extra fermions are introduced in the model. Those extra fermions might not be related to dark matter observables. Since we are studying the $Z^{\prime}$ portal from a general perspective, we will not include extra fermions in our analysis and we will assume that they have negligible impact on the phenomenology associated with the dark matter particle. 
\end{itemize}
\section{Direct Detection}
In direct detection, the relevant observables  are the dark matter-nucleon scattering cross section and the particle dark matter mass. In general, in the low-velocity limit, WIMP-Nucleus scattering can be either spin-independent (SI) or spin-dependent (SD), depending on what sort of effective couplings are involved. As shown in Table \ref{table1}, the $O_3$ and $O_4$ operators induce spin-dependent interactions, whereas $O_1$ and $O_2$ spin independent ones. Therefore the $Z^{\prime}$ portal is subject to both spin-dependent and spin- independent bounds. 

Spin-dependent bounds are generically weaker than Spin Independent, because Spin-independent scattering is proportional to $A^2$ unless a destructive interference happens, which is not the case here \cite{xenonphobia1,xenonphobia2}. As far as direct detection is concerned only the pure vector and vector-axial operators are relevant, because a mixing between these two described by the $O_2$ and $O_3$ operators are velocity suppressed, as shown in Table \ref{table1}. The scattering cross section for the pure vector and vector-axial coupling are given in  Table \ref{table1}. Since spin-dependent bounds provide weaker bounds on the parameter space of the $Z^{\prime}$ portal sort of models we will focus our analysis on the spin-independent case only \cite{ZpComplementary1,ZpComplementary2,ZpComplementary3,ZpComplementary4,ZpComplementary5}. Note that the argument above fails in theories where the vector coupling is extremely suppressed compared to the vector-axial coupling. The vector and axial couplings might be different, as it happens in many models, but they would presumably not differ by orders of magnitude \cite{ZpDM1,ZpDM2,ZpDM3,ZpDM4,ZpDM5,ZpDM6,ZpDM7,ZpDM9,ZpDM10}. 

In any case, hereafter our results regarding collider, indirect detection and direct detection searches are obtained using the Lagrangian given in Eq. (\ref{eq1}) without assuming any particular operator. It is important to notice that despite being a leptophobic theory, the inclusion of leptonic channels would have no impact on the direct detection bounds derived here. The lepton channels are rather relevant for indirect and collider purposes, though \cite{dileptonCMS}. Throughout this work we use the Micromegas package to compute direct detection observables \cite{micromegas1,micromegas2,micromegas3}. 
\section{Indirect Detection}
Indirect searches for dark matter probe different and complementary dark matter observables to direct detection, namely: the annihilation cross section, the dark matter particle mass, and the dark matter halo model. Different halo models have tend to converge to similar trends at distances far enough from the center of given astrophysical objects such as the Galactic center, but they might vary from cuspy to core types at small radial distances \cite{halos1,halos2,halos3}. It is still open to debate whether or not the halo profile of our own Milky Way is cuspy or cored. The Fermi-LAT collaboration \cite{Fermidwarfs}, as well as other independent studies \cite{ID6,danPRL,FarinaldoGC}, have set stringent limits on the annihilation cross section of dark matter particles under different halo profile and annihilation channels assumptions. Here we will use as reference the current Fermi-LAT ones, which are focused on the $b\bar b$ annihilation channel. 

It is important to emphasize that, because we are in the leptophobic regime, the dark matter particle $\chi$ will always annihilate into quarks, which all feature similar $\gamma$-ray spectra, resulting from the neutral pion two-photon decay. We will sum the annihilation modes into all quarks as if they produced the same gamma-ray spectrum of $b\bar b$ case. This approximation is quite reasonable because in fact all quarks produce basically the same gamma-ray spectrum (the possible exception being the top quark near threshold). Once we have summed up the total annihilation cross section into quarks, we compare the result with the current Fermi-LAT bounds, which was obtained from stacked 4-years of observations of local dwarf spheroidal galaxies, assuming an annihilation $100\%$ into $b\bar b$. The Fermi-LAT constraints can be thus straightforwardly applied to our leptophobic setup. 
%
\section{Collider Bounds}

\subsection{Bounds from the search for new resonances in dijet events}

Several extensions of the SM predict the existence of new heavy particles that are within the reach of hadron colliders. A new heavy neutral gauge boson as the $\zp$, for example, may lead to a resonance in $jj$ and $\ell^+\ell^-$ invariant masses, $m_{jj}$ and $m_{\ell\ell}$, respectively. 

The Tevatron and LHC collaborations have been searching for resonances in dijet events with null results until now~\cite{tevjj,lhc7jj,lhc8jj}. These null results allow to place strong constraints on any new model predicting resonances in dijet events, as in our case, where the dark $\zp$ has a large branching fraction into quarks due to leptophobia.

\begin{figure}[t]
\centering
\includegraphics[scale=0.5]{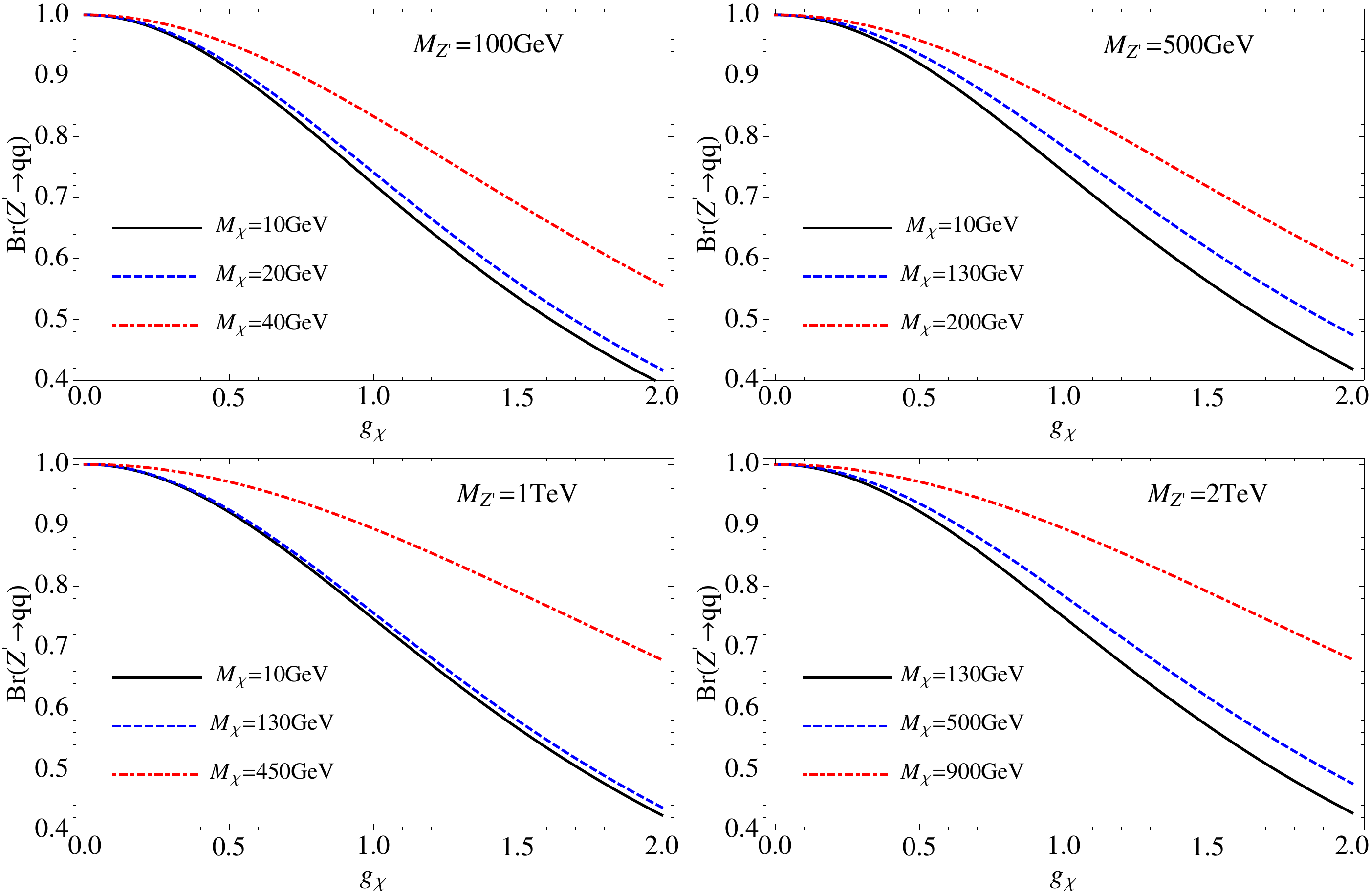}
\caption{The branching ratio $BR(\zp\to q\bar{q})$ as a function of the DM mass $\mchi$ and the $\zp\chi\bar{\chi}$ coupling, $\gx$.}
\label{figbrzpgx}
\end{figure}

Concerning the way a dijet search might constrain a $\zp$ dark matter model, the effect on the mass and couplings of the DM particle is indirect. For a fixed $\zp$ mass, the branching ratio $BR(\zp\to q\bar{q})$ increases as $\mchi$ approaches $\mzp/2$, up until 
\begin{equation}
BR(\zp\to q\bar{q})+BR(\zp\to\chi\bar{\chi})=1.
\end{equation}
Also, decreasing the coupling $\gx$ between the DM and $\zp$ increases the branching faction into quarks. We show in Fig.~\ref{figbrzpgx} the branching ratio into quarks of a $\zp$ of fixed mass as a function of the DM mass $\mchi$ and the $\gx$ coupling.

From Fig.~\ref{figbrzpgx} we see that a dijet search for a $\zp$ is moderately sensitive to the DM model details. For not too large $\gx$ couplings, the branching ratio is never smaller than $\sim 70$\% independent of the DM mass. On the other hand, the $\zp$ mass and couplings to quarks are crucial, not only because of the size of the production cross sections, but also because heavier resonances give rise to harder yields, in this case harder jets which are likely to pass selection cuts.

We now discuss in detail the constraints from Tevatron, LHC 7 and LHC 8 data.

\vskip0.5cm
\noindent{\underline{Tevatron $1.96$ TeV}}
\vskip0.5cm

To evaluate the impact of the Tevatron search on dijet resonances after $1.13\ifb$ of integrated luminosity~\cite{tevjj}, we simulated the process
\begin{equation}
p\bar{p}\to \zp\to jj
\label{ppxjj}
\end{equation}
plus up to two extra jets using \texttt{MadGraph5}~\cite{mad5}--FeynRules~\cite{feynrules}, clustering and hadronizing jets with \texttt{Pythia}~\cite{pythia}, and simulating detector effects with \texttt{PGS4}~\cite{pgs}. Soft and collinear jets from QCD radiation generated by \texttt{Pythia} are consistently merged with the hard radiation calculated from matrix elements in MLM scheme~\cite{mlm} at appropriate matching scales. We adopted the CTEQ6L parton distribution functions computed at $\mu_F=\mu_R=\mzp$. All the collider events simulated in this work were obtained using these packages. 

The signal events were then selected with the same criteria adopted in Ref.~\cite{tevjj} which, by the way, are rather inclusive. The only requirement is on the jets rapidities $y_j$ of an event,
\begin{equation}
|y_j| < 1.
\label{cutev}
\end{equation}
As in the Tevatron analysis of Ref.~\cite{tevjj}, we also multiplied the cross sections by a $K$-factor of 1.3 to account for higher order QCD corrections. We generated events for $\zp$ of masses from $300$ GeV to $1.4$ TeV, the mass region for which the Tevatron data have sensibility, and applied the 95\% C.L. upper limits, quoted in Ref.~\cite{tevjj}, on the production cross section times branching ratio for a $\zp$ after imposing Eq.~(\ref{cutev}).

The exclusion regions, in the $\mzp$ {\it versus} $\gx$, from Tevatron searches, can be seen in Figures~\ref{compare} and \ref{reduce}. The wavy aspect of the curves just reflects the shape of the experimental exclusion regions~\cite{tevjj}.

\vskip0.5cm
\noindent{\underline{LHC 7 TeV}}
\vskip0.5cm

As in the case of the Tevatron, we generated signal events for the process
\begin{equation}
pp\to \zp\to jj
\label{ppjjlhc}
\end{equation}
at the LHC at 7 and 8 TeV center-of-mass energy, plus one and two additional hard jets. 

The ATLAS collaboration performed an early resonance search in dijet invariant masses using $1\ifb$ of data, placing 95\% C.L. upper limits on $\sigma\times BR$~\cite{lhc7jj}. Background suppression was achieved imposing the following cuts on dijets events which we also applied to our signal events,
\begin{eqnarray}
p_{T_j} & > & 180\gev\;\; ,\;\; |\eta_j| < 2.8 \nonumber \\
m_{jj} & > & 717\gev\;\; ,\;\; |y^*|<0.6
\label{cuts7}
\end{eqnarray}
where $y^*=(y_1-y_2)/2$ is the rapidity of the two highest $p_T$ jets in their mutual CM system.

The constraints on the dark $\zp$, in this case, could not be straightforwardly taken from the upper limits quoted in Ref.~\cite{lhc7jj}, once a Gaussian model template, where events are normally distributed in $m_{jj}$, was used to derive the limits, and the jets invariant mass of the heavy $\zp$ resonances are skewed distributions with considerable asymmetries.

In order to obtain the attainable limits from the LHC 7 TeV data we fit the observed dijet invariant mass to the functional form in the 1 to 4 TeV range
\begin{equation}
\frac{d\sigma}{dm_{jj}}=p_0(1-x)^{p_1} x^{p_3+p_4\ln x}
\end{equation}
where $x=m_{jj}/\sqrt{S}$ and the $p_i$ are fit parameters. 

Invariant mass distributions for signal events corresponding to $\zp$ masses from 900 GeV to 3 TeV were generated and the same Bayesian method used in the experimental study to obtain the new 95\% C.L. limits on $\sigma\times BR$. We have assumed a flat prior probability density for the number of signal events and marginalized over a nuisance parameter to account for systematic uncertainties on the signal acceptance. This nuisance parameter was first tuned to reproduce the limits quoted in Ref.~\cite{lhc7jj} for a Gaussian $m_{jj}$ distribution for each $\mzp$. After tunning the systematic uncertainty, we calculated the likelihood function and the posterior probability density function upon which we obtained the new upper limits.

Compared to Tevatron, the LHC 7 TeV exclusion regions are narrower concerning the $\zp$ masses, but it reaches bigger $\gx$ values as can be seen in Figures~\ref{compare}. This is consequence of the much more restrictive selection cuts of LHC 7 TeV analysis necessary to suppress the QCD backgrounds, and larger production cross sections compared to Tevatron. Next we comment the dijet production at the LHC 8 TeV. 

\vskip1.5cm
\noindent{\underline{LHC 8 TeV}}
\vskip0.5cm

The CMS collaboration performed a search for narrow resonances using dijets with $19.6\ifb$ of data at the LHC 8 TeV~\cite{lhc8jj}. Contrary to the ATLAS search~\cite{lhc7jj}, 95\% C.L. limits on a $\zp$ model were derived in the 1 to 5 TeV range. In this case, all jets in the analysis were requested to have
\begin{equation}
p_{T_j} > 30\gev\;\; ,\;\; |\eta_j| < 2.5
\label{cuts8}
\end{equation}
and if one of two highest $p_T$ (leading) jets fails to pass these cuts, the event is discarded.

To reduce the sensitivity to gluon radiation, the remaining jets are combined into {\it wide jets}~\cite{wide}, $J$, which are constructed from the leading jets, $j_{1,2}$, in an event by adding the four-vectors of all other jets to the closest leading jet if $\Delta R_{ij}=\sqrt{\left(\Delta R_{ij}\right)^2+\left(\Delta \phi_{ij}\right)^2}<R_w$, $i=1,2$ and $j$ denotes a non-leading jet. As in the experimental study we set $R_w=1.1$ to form {\it wide jets} in signal events. Now, the dijet system is composed of two {\it wide jets} upon which we impose the following additional cuts
\begin{equation}
|\Delta\eta_{JJ}| < 1.3 \;\; ,\;\; m_{JJ} > 890\gev.
\label{widecuts}
\end{equation}
\begin{figure}[t]
\centering
\includegraphics[scale=0.4]{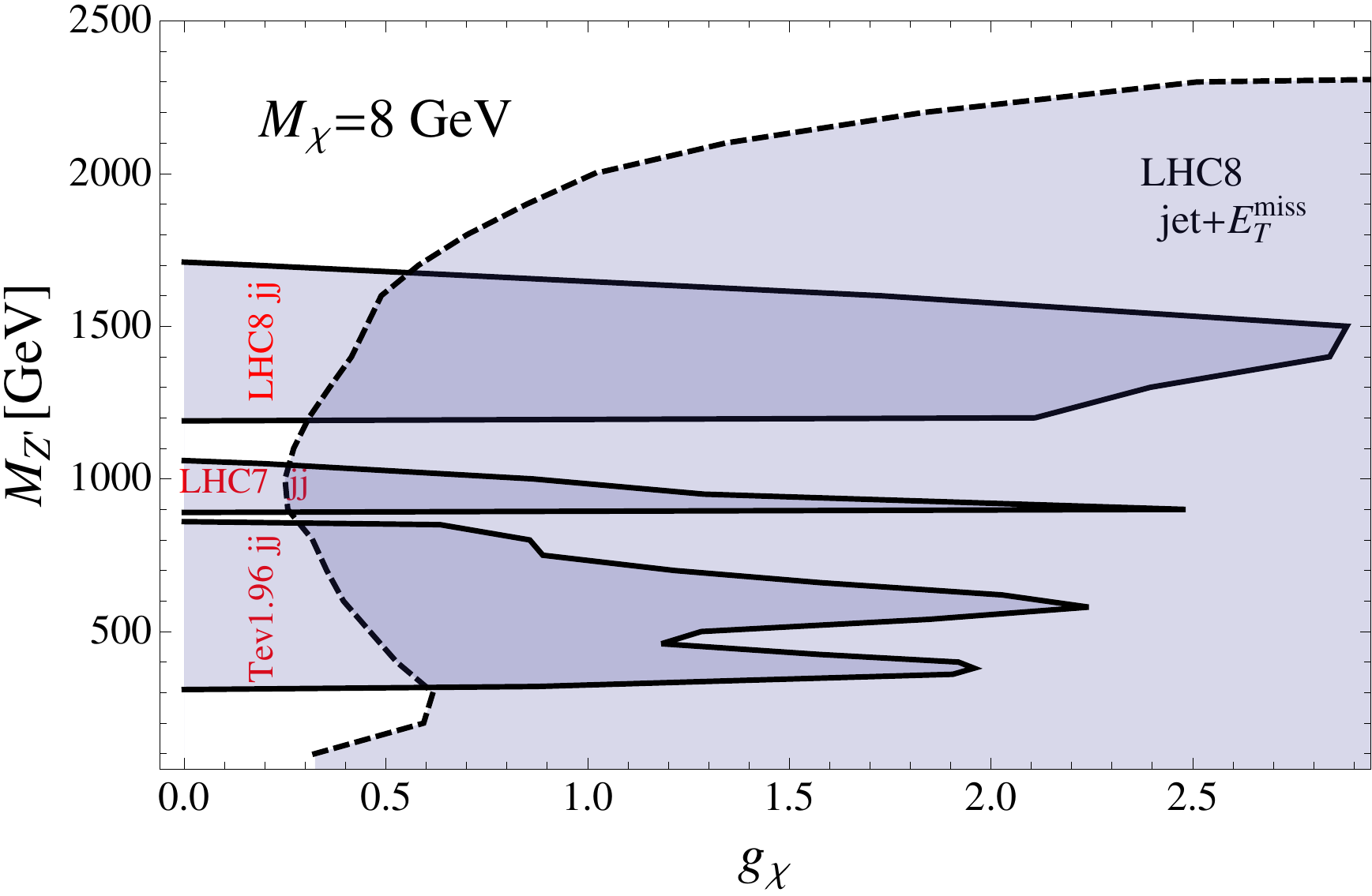}
\includegraphics[scale=0.4]{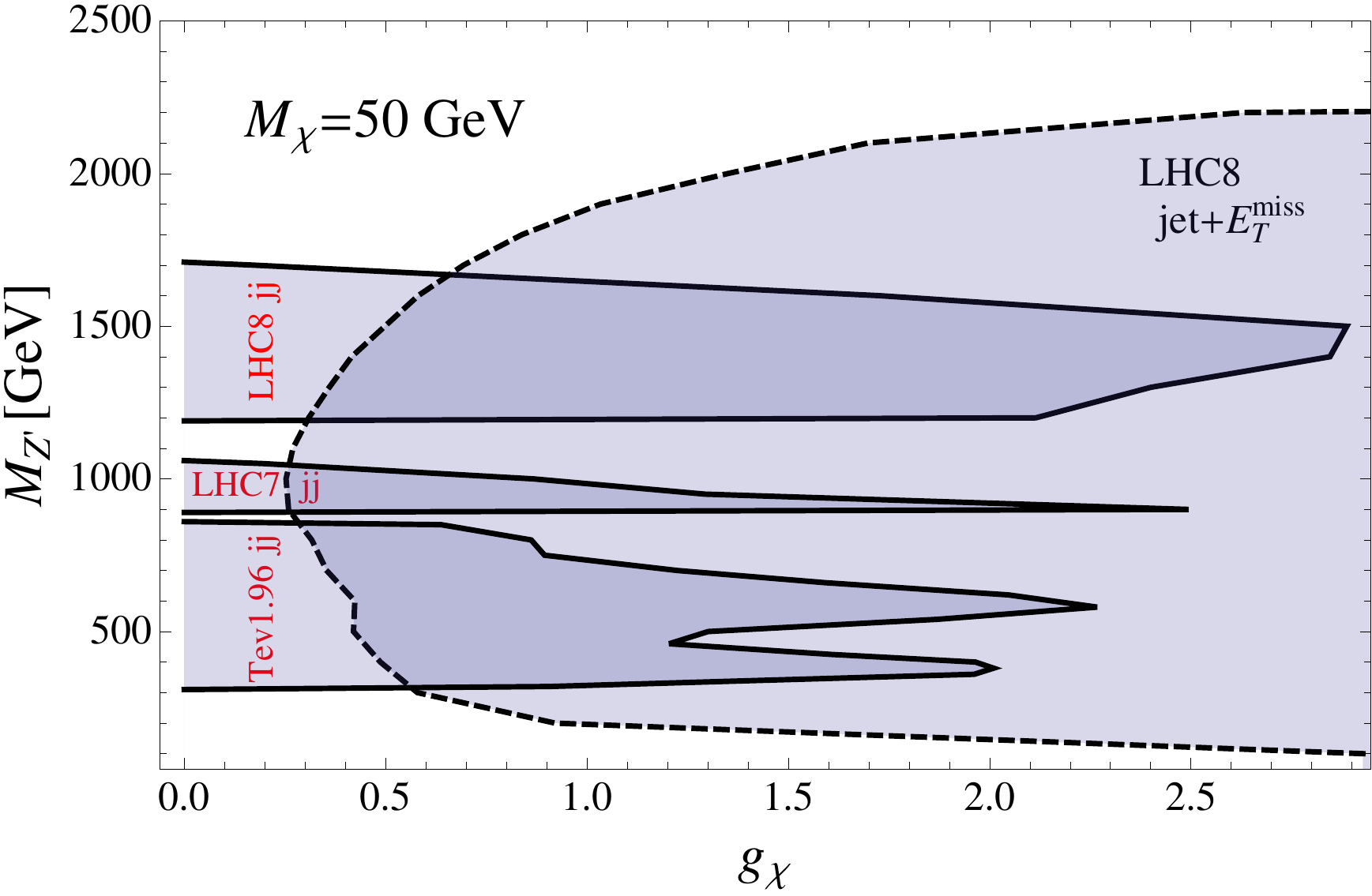} \\
\includegraphics[scale=0.4]{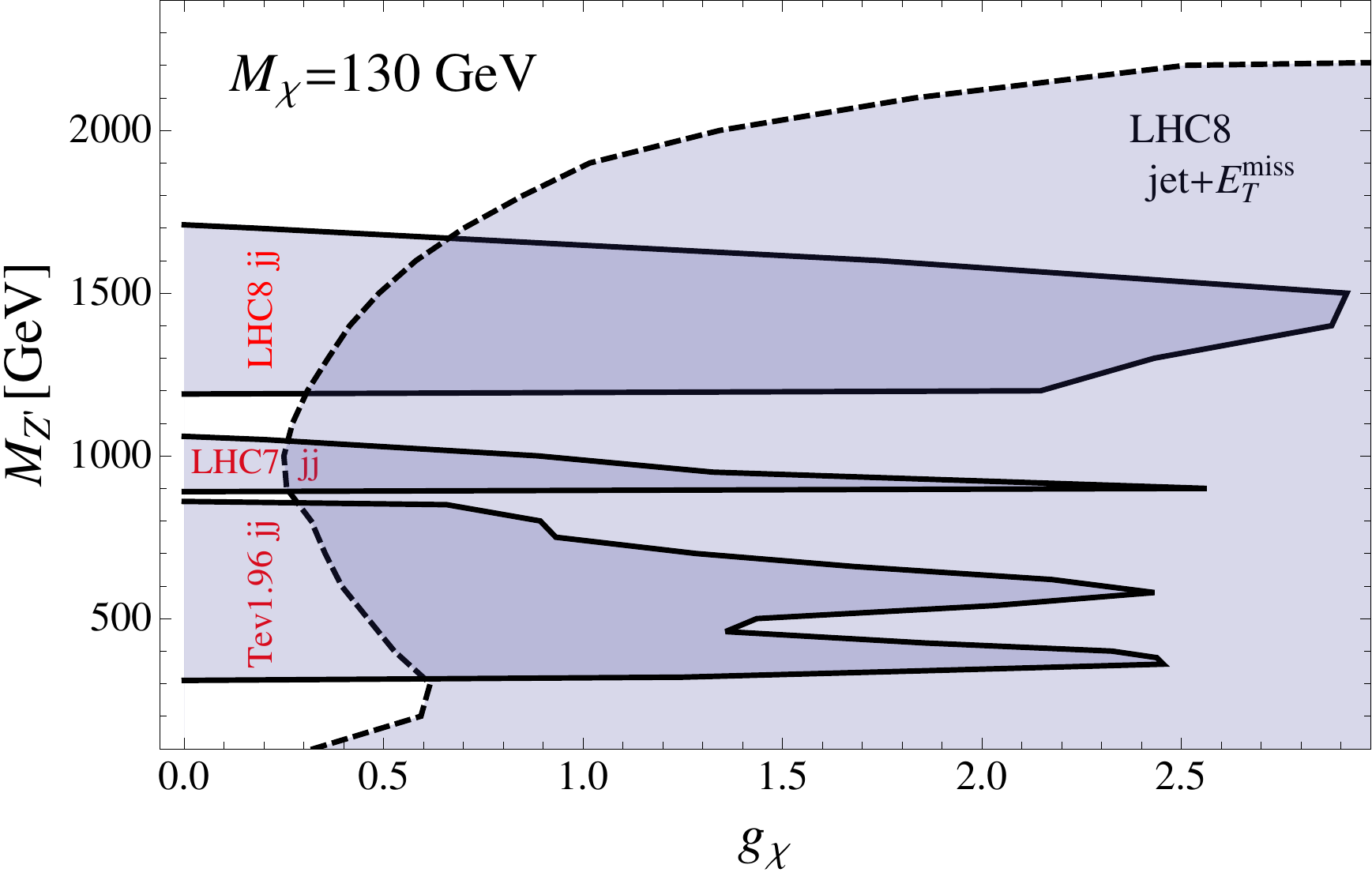}
\includegraphics[scale=0.4]{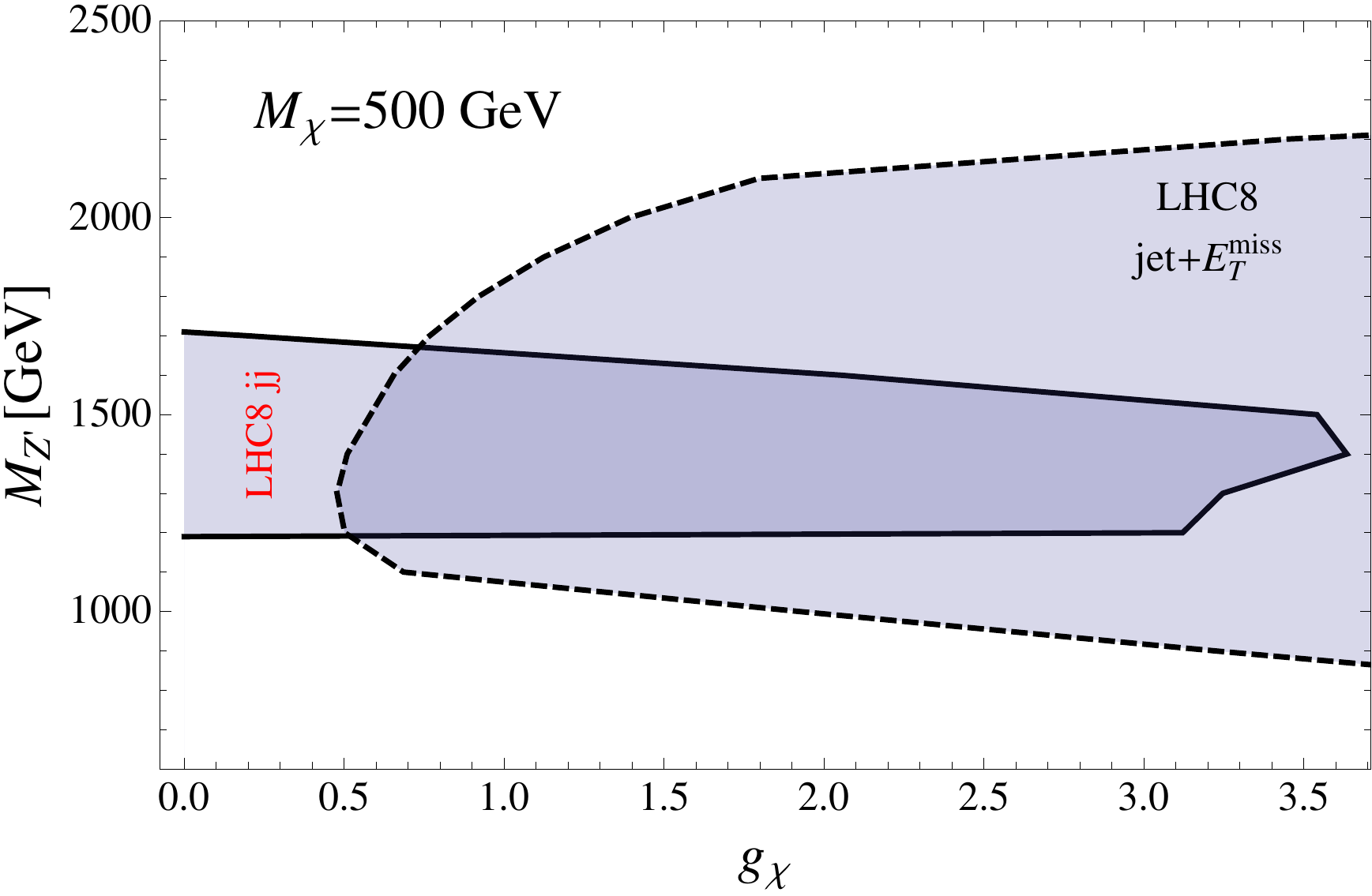}
\caption{Exclusion regions from Tevatron, LHC 7 and 8 TeV searches for resonances in dijet events and from monojet search at the LHC 8 TeV in the $\mzp$ {\it versus} $\gx$ plane. The left upper panel shows the 95\% C.L. excluded region for a fixed DM mass of 8 GeV, while the upper right, the lower left, and lower right panels show the cases for a 50, 130, and 500 GeV, respectively.}
\label{compare}
\end{figure}

Compared to the Tevatron and LHC 7 TeV searches, the CMS LHC 8 TeV analysis is able to exclude bigger $\gx$ couplings for a given DM mass  in the $\sim 1.2$ to $1.7$ TeV mass range as can be seen in Fig.~\ref{compare} for DM masses from 8 to 500 GeV. 

The LHC 7 TeV has a deeper reach compared to the Tevatron, but in a narrower mass range. This is consequence of a more selective set of kinematic cuts, as we discussed. The LHC 8 TeV, on the other hand, excludes a larger region of the $\mzp$ {\it versus} $\gx$ plane not only because of the larger production cross sections, but also due to the more efficient selection of signal events and the much larger integrated luminosity as compared to previous studies.

Note, however, that there exist small gaps between the $\mzp$ regions covered by the three sets of data, mainly between LHC 7 TeV and LHC 8 TeV. All those gaps could be closed using the whole of data accumulated by the Tevatron and 7 TeV run of the LHC. This notwithstanding, there is an almost complete complementarity between the 3 experiments for small $g_x$ couplings. It should also be pointed out the complementarity between dijet and monojet searches as we are going to discuss next.

\subsection{Bounds from DM searches in the monojet channel}

The associated production of DM and jets, photons or gauge bosons, has been extensively studied both from the theoretical and experimental sides in the search for DM in colliders.

In models with a dark mediator, as the $\zp$, the process we are interested in is
\begin{equation}
pp\to \zp+j\to \chi\bar{\chi}+j\to \eslash +j
\end{equation}
where a jet from QCD radiation is irradiated from the initial state partons alongside $\zp$ which, then, decays to a pair of DM particles giving rise to a hard jet and large missing energy signal.

As in the case of dijets, we simulated events with one additional jet and matching the hard matrix element contributions from \texttt{MadGraph5} to soft and collinear jets from \texttt{Pythia} at an appropriate matching scale in the MLM scheme. The renormalization and factorization scales were chosen dynamically in this case as $\mu_R=\mu_F=\sqrt{p_T^2+\frac{1}{2}\mzp^2}$ where $p_T$ is the transverse momentum of the hardest jet in the event.
\begin{figure}[t]
\centering
\includegraphics[scale=0.5]{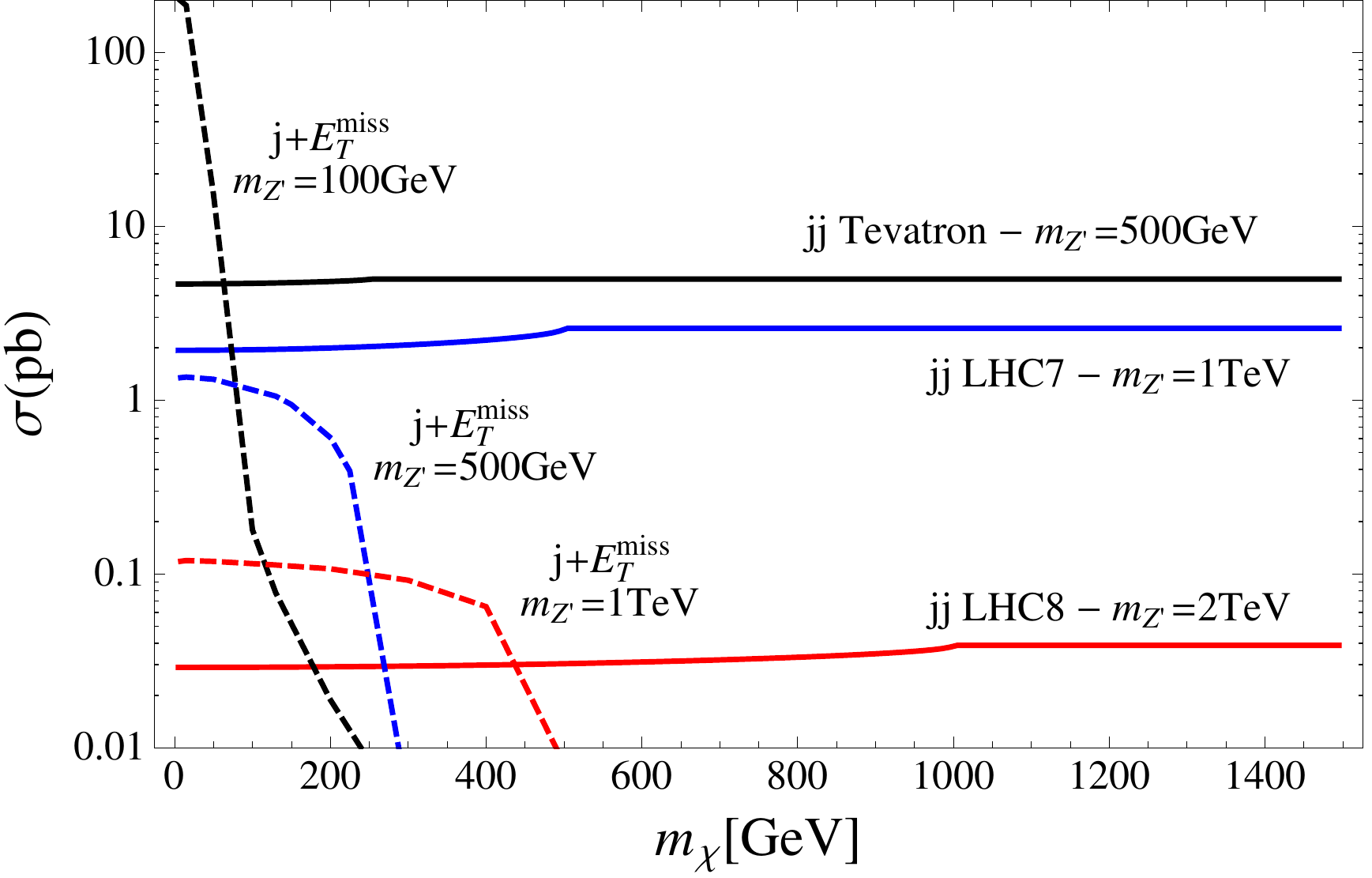}
\caption{Production cross sections for dijet (solid lines) and monojet (dashed lines) processes as a function of the DM mass for three $\zp$ masses at the Tevatron $1.96$ TeV, LHC 7 and 8 TeV.}
\label{xsec}
\end{figure}

Combining dijet and monojet searches for DM is interesting once they cover complementary regions of the $\mzp$ {\it versus} $\gx$ plane. As we discussed, dijets have a larger sensitivity to small $\gx$ couplings, as a larger $\zp\chi\bar{\chi}$ coupling increases the branching ratio do DM. On the other hand, monojets have an increased production rate for larger $\gx$ once $\zp$ decays to DM in this case. Also, the sensitivity to the DM mass is complementary between the two processes. As the DM gets heavier, the $BR(\zp\to\chi\bar{\chi})$ decreases for a fixed $\mzp$, which enhances the dijet rates, but suppresses the number of monojet events. This effect can be seen in Figure~\ref{xsec} where we show the production rates as a function of the DM mass.

The CMS collaboration has performed a search for new physics in monojet events with $19.5\ifb$ of data at the 8 TeV LHC~\cite{mono8}, placing 95\% C.L. upper limits on new physics events passing the following selection criteria
\begin{eqnarray}
p_{T_{j_1}} &>& 110\gev \;\; ,\;\; |\eta_{j_1}| < 2.4 \nonumber \\
\eslash &>& E_T^{miss}
\end{eqnarray}
where $p_{T_{j_1}}$ is the transverse momentum of the leading jet of the event. A second jet is allowed provided it is not too far away from the leading jet in azimuthal direction, $\Delta\phi(j_1,j_2)<2.5$. Events with more than two jets with $p_T > 30\gev$ and $|\eta|<4.5$ are vetoed.

Seven $E_T^{miss}$ regions were used to select signal events: $250$, $300$, $350$, $400$, $450$, $500$, and $550\gev$. Requiring the signal events to satisfy these criteria excludes adjacent and intersecting regions of the $\mzp$ {\it versus} $\gx$ plane. The resulting 95\% C.L. exclusion region is shown in Fig.~\ref{compare} in the case of an 8, 50, 130 and 500 GeV DM mass.
 
It should be pointed out how the complementarity between dijet and monojet searches excludes entire regions of the parameters space. Except for small gaps, all $\zp$ masses from 300 GeV to 1.7 TeV are excluded for all $\zp$
 couplings to DM at 95\% confidence level in the case of an 8 GeV DM mass, for example, as is shown in the upper left panel of Figure~\ref{compare}. In fact, very similar regions are excluded for DM masses up to 130 GeV, at least, as can be seen in Figure~\ref{compare}. Even for a 500 GeV DM, $\zp$ masses fro 1.2 to 1.6 TeV are excluded at 95\% C.L. for all $\gx$ as we see in the lower right panel of Fig.~\ref{compare}.

Some models predicting a new heavy gauge boson, such as 331 models, and other models with extended gauge groups, might present reduced gauge couplings between quarks and the new gauge bosons. Reducing the $\zp-q-q$ couplings, $g_{\zp qq}$, shrink all exclusion regions collectively, and both dijet and monojet data are more easily evaded. We show the effect of reducing $g_{\zp qq}$ by 90\% and 50\% for a 130 GeV DM in Fig.~\ref{reduce}.
\begin{figure}[t]
\centering
\includegraphics[scale=0.4]{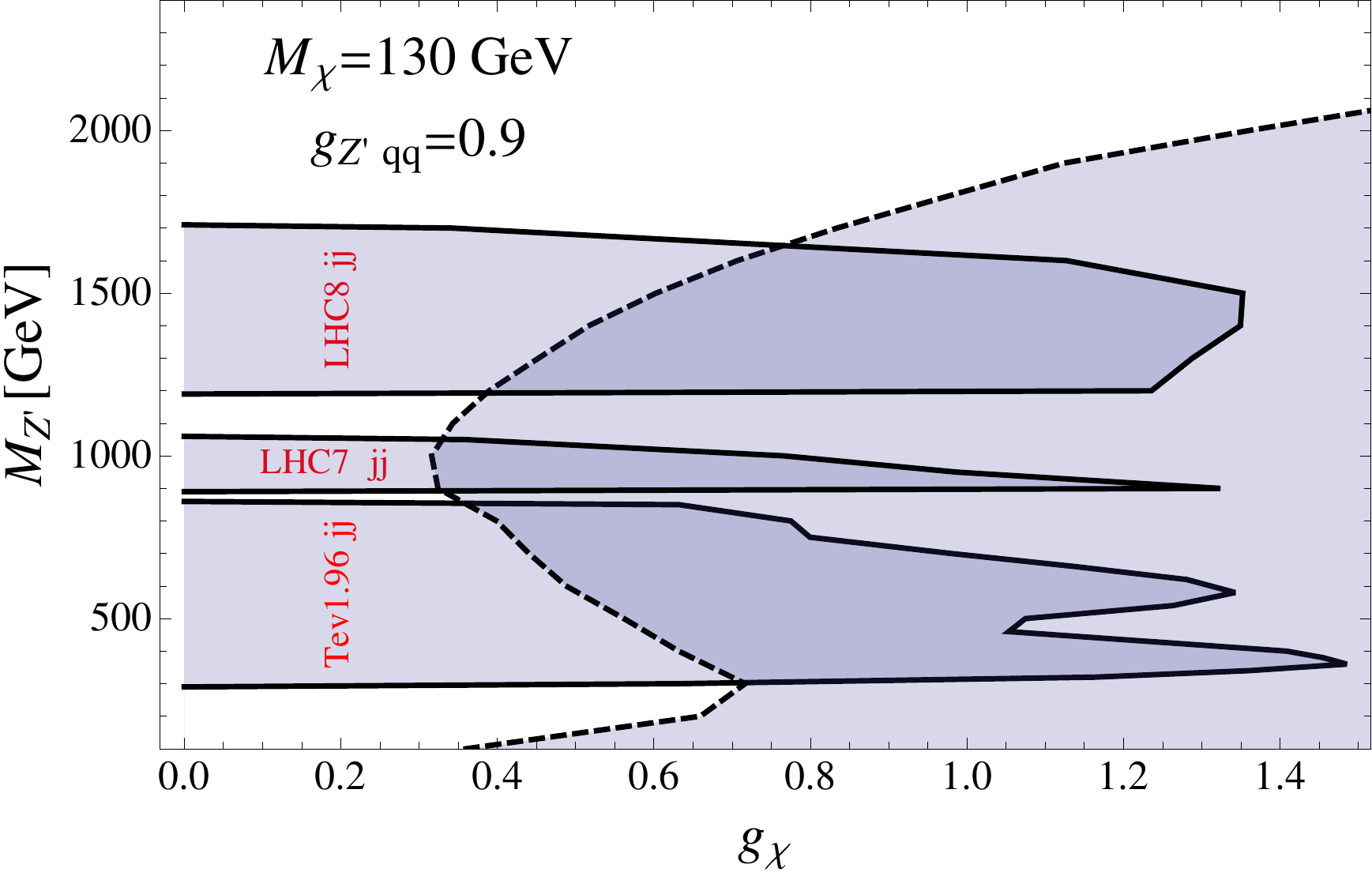}
\includegraphics[scale=0.4]{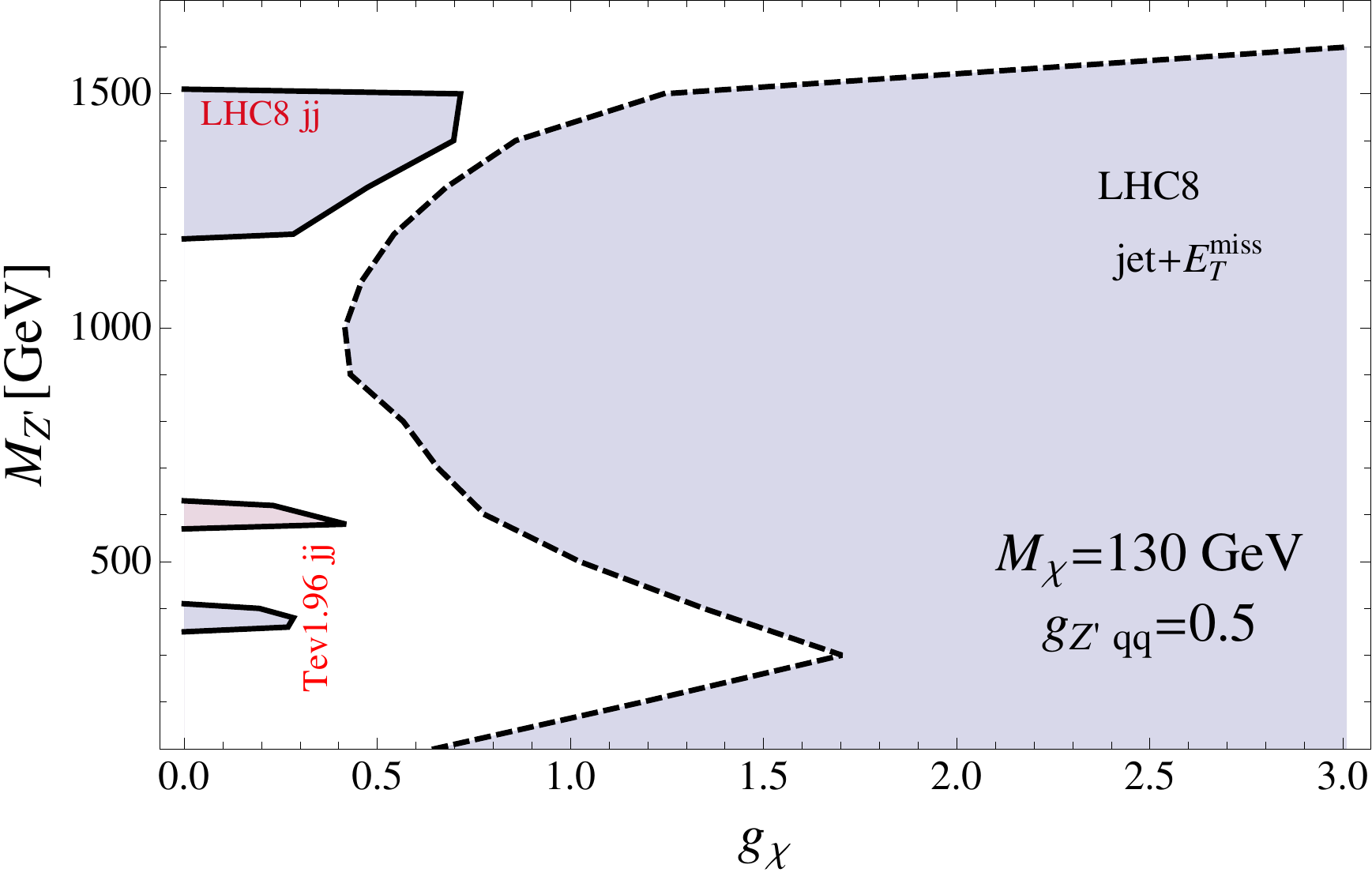}
\caption{Exclusion regions from Tevatron, LHC 7 and 8 TeV searches for resonances in dijet events and from monojet search at the LHC 8 TeV in the $\mzp$ {\it versus} $\gx$ plane. The left panel shows the 95\% C.L. excluded region for a fixed DM mass of 130 GeV and a 90\% diluted coupling between $\zp$ and quarks, and the right panel the same but for a 50\% diluted coupling.}
\label{reduce}
\end{figure}

We now discuss the complementarity among direct, indirect and collider searches of DM and highlight their interplay to constrain a dark $\zp$ model such as the one considered in this work.

\section{Complementary Results: Direct, Indirect Detection and Collider}
So far we have discussed direct, indirect and collider searches for DM in the particular $\zp$ portal scenario under consideration. We here investigate how the parameter space of the  $Z^{\prime}$ portal is constrained by these bounds. We will point out how crucial complementarity is when it comes to dark matter searches. Direct detection bounds such as the XENON100/LUX and the LHC 8 TeV jet+$E_T^{miss}$ limits, are both more sensitive to larger $\gx$ ($DM-DM-Z^{\prime}$) couplings. Indirect detection constraints coming from Fermi-LAT Dwarf Spheroidals data rule out light $Z^{\prime}$ masses and a wide range of $\gx$ couplings.  Other collider bounds coming from the LHC 7 TeV and Tevatron $1.96$ TeV dijet data give complementary bounds on small $g_{\chi}$ couplings for larger $Z^{\prime}$ masses. We select WIMP masses based on the tantalizing signals reported in both direct and indirect searches, although we will not attempt to fit for those signals with the model under investigation.
\subsection{$8$~GeV WIMP in Light of DAMA,CDMS-Si,CoGeNT and CRESST}
An $8$~GeV WIMP is a well motivated dark matter candidate because of recent positive signals coming from direct detection. In particular, the CoGeNT collaboration has claimed to observe a $\sim 2\sigma$ modulation consistent with a $8$~GeV WIMP scattering off nuclei with a spin independent cross section of $3-4\times 10^{-5}$~pb \cite{DDexp2,DDexp3}. Moreover, the  DAMA experiment has observed with $\sim 9 \sigma$ an annual modulation also consistent with a WIMP scattering  \cite{DDexp1}. Furthermore, the three excess events reported by CDMS-Si could also be plausibly explained by a similar WIMP \cite{DDexp5}. For this reason, it is interesting to ask whether a $8$~GeV WIMP via the $Z^{\prime}$ portal is a feasible explanation to these signals, bearing in mind the mentioned LUX and XENON100 constraints.

First, we show in Fig.\ref{figCS1} the spin independent cross section as a function of the $Z^{\prime}$ mass with $g_{\chi}$ varying randomly from zero to one, where the blue (green) points correspond to the underabundant (overabundant) regimes. In the {\it left panel} we have assumed the coupling factors $a=b=1$. Thus we are in the regime that the $Z^{\prime}$ boson couples equivalently to the SM Z. While in the {\it right panel} we have used a $50\%$ suppression in the $Z^{\prime}$-quarks couplings, i.e, with $a=b=0.5$. We thus conclude that in both cases the $Z^{\prime}$ portal with a $8$~GeV DM particle is excluded by the XENON and LUX constraints in the sense that the region of the parameter space that sets the right abundance (thin line between green and blue points) is ruled out by orders of magnitude. In other words, $Z^{\prime}$ mediated processes do not offer the necessary ``XENONPHOBIA'' described in \cite{xenonphobia1,xenonphobia2}, and therefore do not provide a viable explanation for these modulation signals, while still being consistent with other searches. This conclusion might be different if the $Z^{\prime}$ portal is not responsible for setting the WIMP abundance. 

With this in mind, in Fig.\ref{fig1} we show the importance of the complementary search for bounds on the $Z^{\prime}$ portal.  Assuming that the connection between the visible and the dark sector is given exclusively through the $Z^{\prime}$ portal with the coupling factors $a=b=1$ we have computed the abundance and the black line in Fig.\ref{fig1} reproduces $\Omega h^2 = 0.12$. The region beneath the black line provides $\Omega h^2 < 0.12$, whereas the one above the line $\Omega h^2 > 0.12$.

\begin{figure}[!t]
\centering
\includegraphics[scale=0.55]{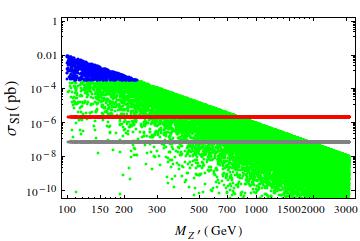}
\includegraphics[scale=0.55]{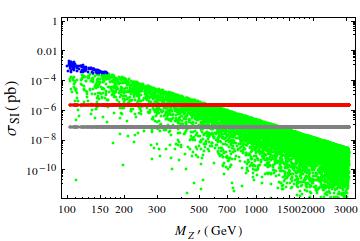}
\caption{Spin independent cross section as a function of the $Z^{\prime}$ for $\mchi=$~8GeV with $a=b=1 (left)$ and $a=b=0.5 (right)$.  Green points provide $\Omega h^2 > 0.12$ (overabundant), whereas the blue ones $\Omega h^2 < 0.11$ (underabundant). The horizontal lines are the XENON100 (top) and LUX (bottom) limits.}
\label{figCS1}
\end{figure}

In Fig.\ref{fig1} we also exhibit the parameter space $M_{Z^{\prime}} \times g_{\chi}$ along with the collider, direct and indirect detection bounds.  The white gaps are not ruled out by any constraint. Pink and Grey regions are ruled out by the LHC $8$~TeV  and $7$~TeV using the dijet data. Yellow is excluded by Tevatron ($1.96$~TeV) with the dijet data. The dashed blue region is excluded by LHC at $8$~TeV using the $jet+ E_T^{miss}$ analysis. The red (blue) dashed region is the XENON100 (LUX 2013) excluded region. We can conclude from Fig.\ref{fig1} that with the recent LUX results the light mediator region is excluded only the mass range $1200 < \mzp < 1100$ with $g_{\chi} < 0.25$ we can evade all constraints. Apart from these, only heavy mediators $\mzp> 1.7$~TeV, are allowed by the current data.

\begin{figure}[!t]
\centering
\includegraphics[scale=0.7]{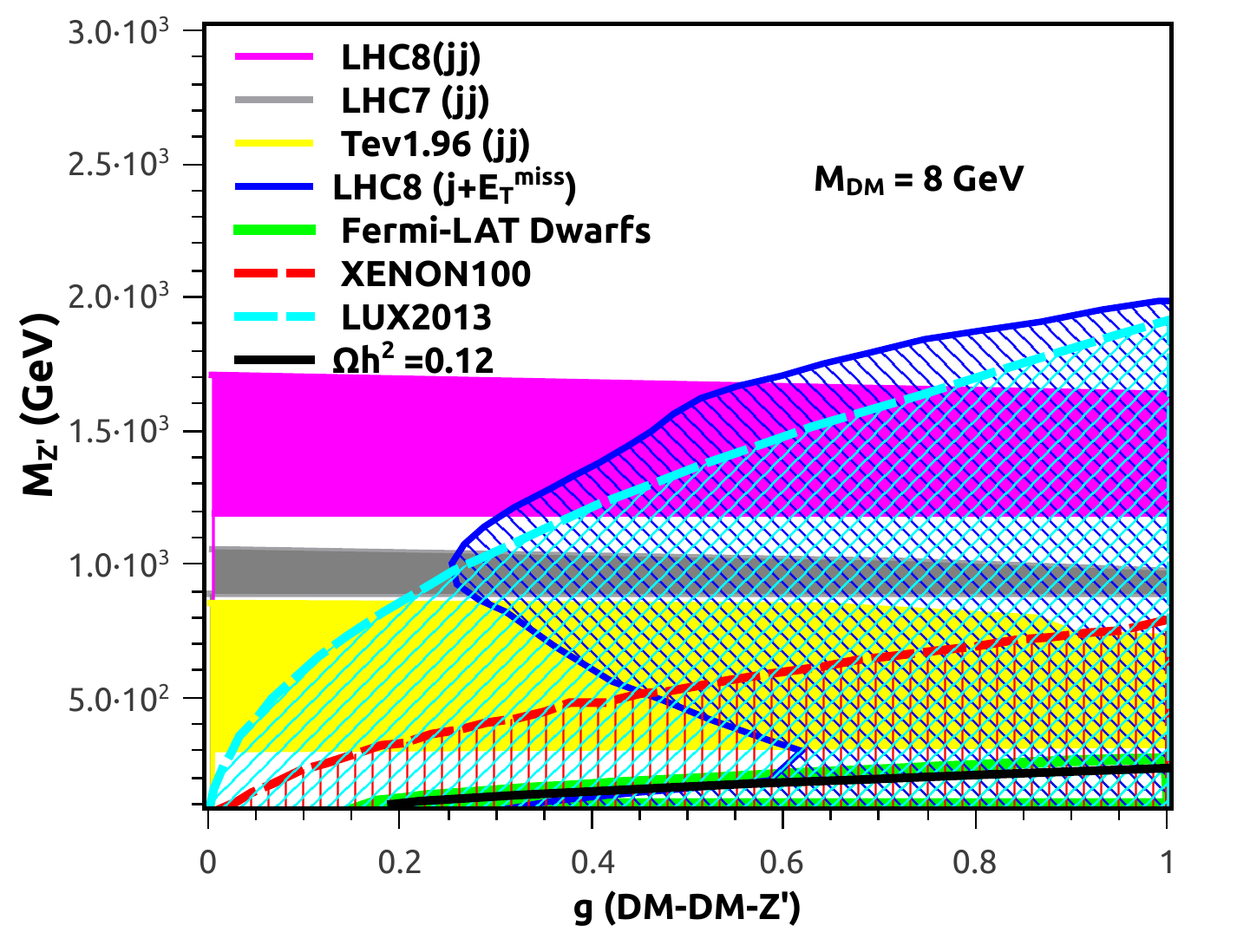}
\caption{Result for $\mchi=8$~GeV in the $M_{Z^{\prime}} \times$ DM-DM-$Z^{\prime}$ coupling plane. The white regions are not ruled out by any constraint. Pink and Grey regions are ruled out by the LHC $8$~TeV  and $7$~TeV using the jet+jet data. Yellow is excluded by Tevatron ($1.96$~TeV) with the jet+jet data. The dashed blue region is excluded by LHC at $8$~TeV using the $jet+ E_T^{miss}$ analysis. The red (blue) dashed region is the XENON100 (LUX 2013) excluded region. The green area in the bottom is ruled out by Fermi-LAT Dwarf Spheroidals bounds. The black line reproduces $\Omega h^2 = 0.12$, whereas the regions below and beneath the lines set $\Omega h^2 > 0.12$ and $\Omega h^2 < 0.12$ respectively.}
\label{fig1}
\end{figure}

In order to probe different $Z^{\prime}$ portal particle physics models we changed the $Z^{\prime}$-quarks coupling by $50\%$, i.e, with the coupling factors determined in Eq.\ref{eq1} as $a=b=0.5$. Again we have calculated the abundance in this regime and drawn a black line in Fig.\ref{fig2} which reproduces $\Omega h^2 = 0.12$. The region under (above) the black line provides $\Omega h^2 < 0.12$ ($\Omega h^2 > 0.12$). It is important to emphasize that we have assumed that the connection between the visible and the dark sector is given exclusively through the $Z^{\prime}$ portal with the coupling factors $a=b=0.5$.

Furthermore, in Fig.\ref{fig2} we exhibit the parameter space $M_{Z^{\prime}} \times g_{\chi}$ along with the collider, direct and indirect detection bounds. The white gaps are not ruled out by any constraint. Pink region is ruled out by the LHC $8$~TeV dijet data. Yellow is excluded by Tevatron ($1.96$~TeV) with the dijet data. The dashed blue region is excluded by LHC at $8$~TeV using the $jet+ E_T^{miss}$ data. The red (blue) dashed region is the XENON100 (LUX 2013)  excluded region, whereas the green region is ruled out be Fermi-LAT Dwarfs. We note how significant the impact of suppressing the $Z^{\prime}$-quarks couplings by $50\%$ is. It opens up a large region of the parameter space consistent with the current bounds. Therefore light and heavy $Z^{\prime}$ bosons with suppressed couplings with quarks are totally a viable annihilation channel as long as they do not set the abundance of the dark matter particle. As we can see in Fig.\ref{fig2}, the black line that delimits the right abundance parameter space lies in the very light mediators region only, but such light mediators are excluded by both current XENON100/LUX and Fermi-LAT Dwarfs bounds. 
\begin{figure}[!t]
\centering
\includegraphics[scale=0.7]{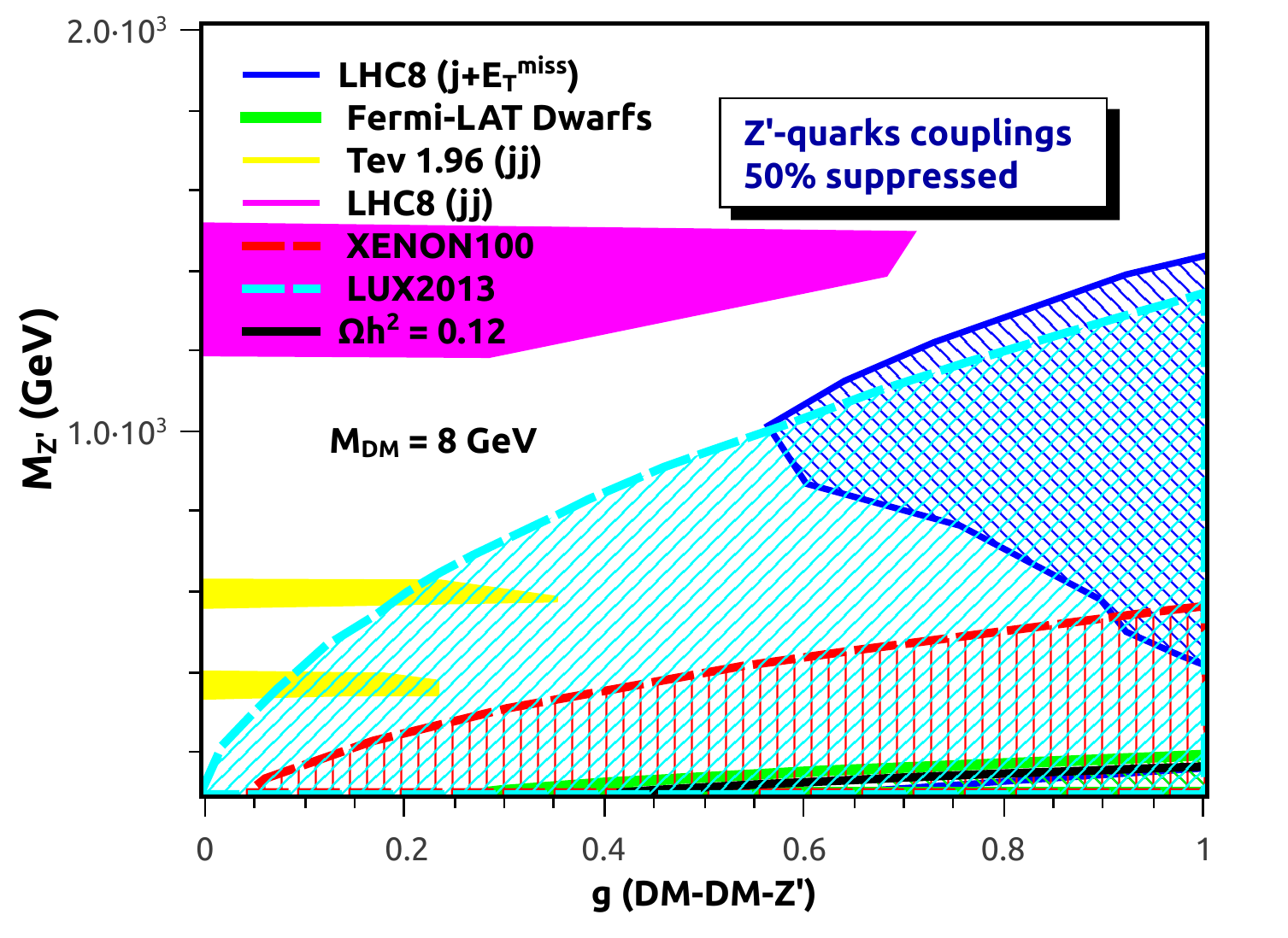}
\caption{Result for $\mchi=8$~GeV in the $M_{Z^{\prime}} \times$ DM-DM-$Z^{\prime}$ coupling plane, with $Z^{\prime}$-quarks couplings suppressed. The white regions are not ruled out by any constraint. The pink region is ruled out by the LHC $8$~TeV  using the dijet data. Yellow is excluded by Tevatron ($1.96$ TeV) with the dijet data. The dashed blue region is excluded by LHC at $8$~TeV using the $jet+ E_T^{miss}$ analysis. The red (blue) dashed region is the XENON100 (LUX 2013) excluded region. The green area in the bottom is ruled out by Fermi-LAT Dwarf Spheroidals bounds. The black line reproduces $\Omega h^2 = 0.12$, whereas the regions below and beneath the lines set $\Omega h^2 > 0.12$ and $\Omega h^2 < 0.12$ respectively.}
\label{fig2}
\end{figure}

\subsection{$15$~GeV WIMP in Light of CDMS-Si Excess Events}
A 15 GeV WIMP is also well motivated because of the three recent excess events observed by CDMS-Si. Despite a likelihood analysis favors a $8$~GeV WIMP, a $15$~GeV DM particle with a SI cross section of $2\times 10^{-6}$~pb is perfectly capable of explaining the signal. Besides, the excess of gamma-ray emission from the Galactic Center which might as well be partially and plausibly explained by a $\sim 15$~GeV WIMP that annihilates mostly in bb \cite{ID1}. For these reasons we will investigate the $Z^{\prime}$ portal for $\mchi=15$~GeV.

First, we show in Fig.\ref{figCS2} the spin independent cross section as a function of the $Z^{\prime}$ mass with $g_{\chi}$ varying randomly from zero to one. In the {\it left panel} we have assumed the coupling factors $a=b=1$, while in the in the {\it right panel} we have used a $50\%$ suppression in the $Z^{\prime}$-quarks couplings, i.e, with $a=b=0.5$. We again conclude that in both cases the $Z^{\prime}$ portal with $15$~GeV dark matter particle is excluded by the XENON100 (top) and LUX (bottom) bounds. Therefore a $15$~GeV DM particle which scatters off nuclei via the $Z^{\prime}$ boson is not a viable explanation for this excess events while obeying other constraints. One would need more exotic gauge coupling structures to acquire the ``XENONPHOBIA'' necessary to evade the XENON and LUX bounds. 

In Fig.\ref{fig3} draw a black line which delimits the right abundance ($\Omega h^2 = 0.12$) parameter space for a $15$~GeV DM particle as a function of the DM-DM-$Z^{\prime}$ coupling ($g_{\chi}$) assuming that the connection between the visible and the dark sector is given through the $Z^{\prime}$ portal only with the coupling factors $a=b=1$, i.e, in the regime that the $Z^{\prime}$ boson has identical couplings to the SM Z.  The region on top of the black line provides $\Omega h^2 > 0.12$, whereas the beneath the line one set $\Omega h^2 < 0.11$. Hence $Z^{\prime}$ mediated processes overproduce the dark matter particles unless we are in the regime of light mediator $M_{Z^{\prime}} < 500$~GeV.

In Fig.\ref{fig3} we show as well the parameter space $M_{Z^{\prime}} \times g_{\chi}$ along with the collider, direct and indirect detection bounds.  The white regions are not ruled out by any constraint. Pink and Grey regions are ruled out by the LHC $15$~TeV  and $7$~TeV using the dijet data. Yellow is excluded by Tevatron ($1.96$~TeV) with the dijet data. The dashed blue region is excluded by LHC at $15$~TeV using the $jet+ E_T^{miss}$ analysis. The red (blue) dashed region is the XENON100 (LUX 2013) excluded region, whereas the green one is excluded by Fermi-LAT Dwarfs data \cite{Fermidwarfs}. We thus conclude from Fig.\ref{fig3} that the light mediator window is now closed and the region with $M_{Z^{\prime}} \sim 1100$ for $g_{\chi} < 0.05$ is still viable. Apart from this one, only heavy mediators $M_{Z^{\prime}}> 1.7$~TeV with $g_{\chi} < 0.4$, are allowed by the current data. Notice that the light mediator regime is still totally ruled out XENON/LUX bounds in comparison with the $8$~GeV DM case. This can be explained simply by the energy threshold of XENON/LUX. At sufficiently low energies the XENON/LUX efficiency goes down, and therefore for very light dark matter particle $\mchi < 8$~GeV the XENON/LUX bounds gets weaker. On the order hand, as we increase the mass of the DM particle up to $\sim 15$~GeV, XENON/LUX constraints kick in. We clearly see this effect by comparing the XENON100/LUX excluded region in the left panels of Fig.\ref{fig1} and Fig.\ref{fig3}.
\begin{figure}[!t]
\centering
\includegraphics[scale=0.55]{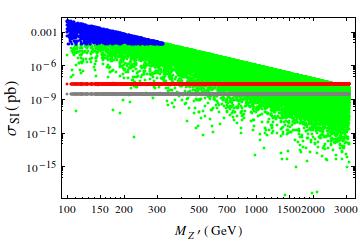}
\includegraphics[scale=0.55]{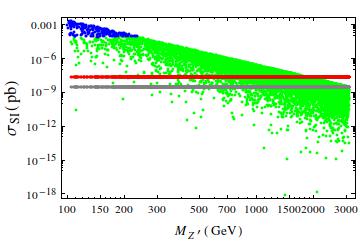}
\caption{Spin independent cross section as a function of the $Z^{\prime}$ for $\mchi=15$~GeV with $a=b=1 (left)$ and $a=b=0.5 (right)$. Green points provide $\Omega h^2 > 0.12$ (overabundant), whereas the blue ones $\Omega h^2 < 0.11$ (underabundant). The horizontal lines are the XENON100 (top) and LUX (bottom) limits.}
\label{figCS2}
\end{figure}
\begin{figure}[!t]
\centering
\includegraphics[scale=0.7]{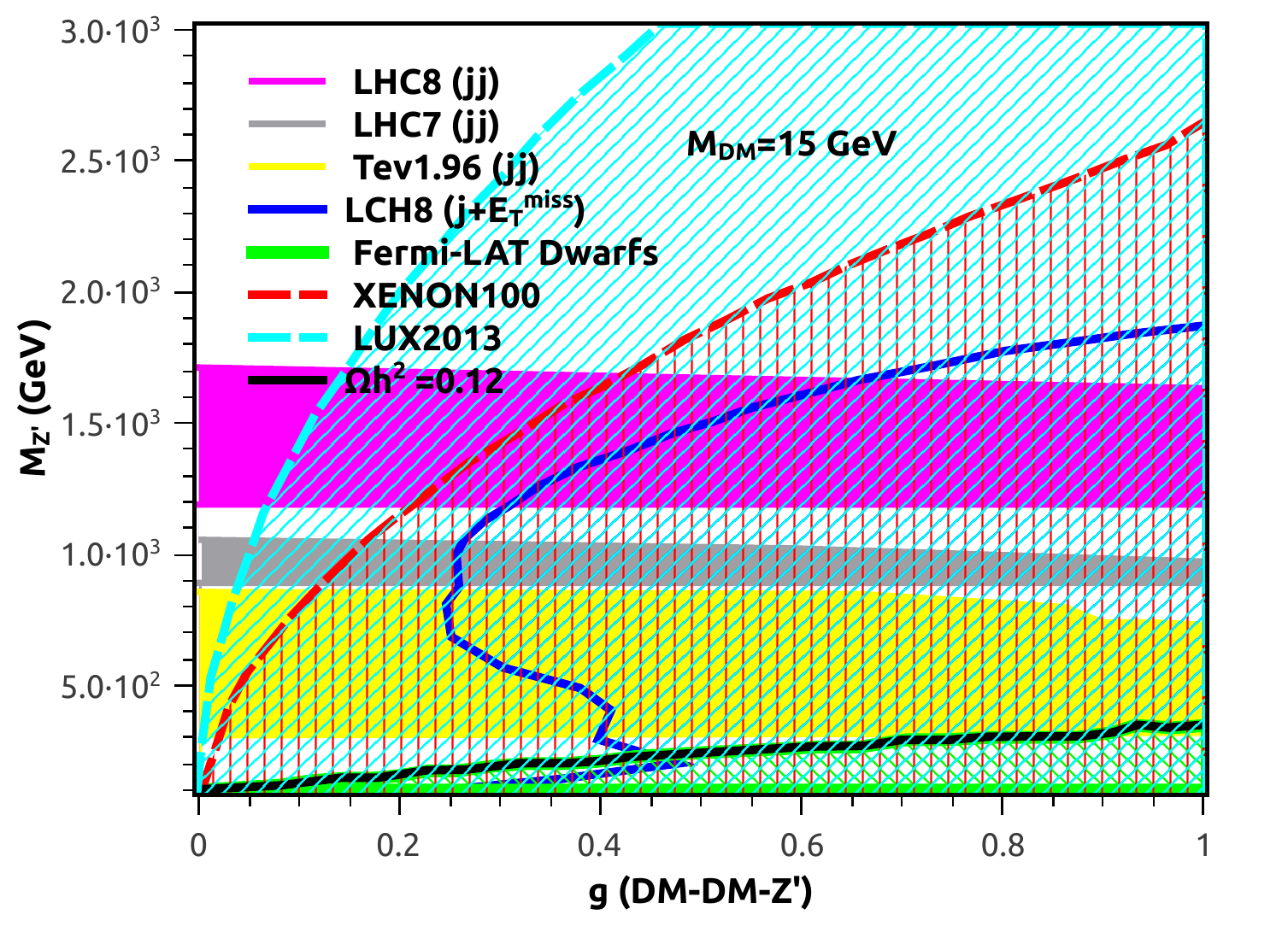}
\caption{Result for $\mchi=15$~GeV in the $M_{Z^{\prime}} \times$ DM-DM-$Z^{\prime}$ coupling plane. The white regions are not ruled out by any constraint. Pink and Grey regions are ruled out by the LHC $8$~TeV  and $7$~TeV using the dijet data. Yellow is excluded by Tevatron ($1.96$ TeV) with the dijet data. The dashed blue region is excluded by LHC at $8$~TeV using the $jet+ E_T^{miss}$ analysis. The red (blue) dashed region is the XENON100 (LUX 2013) excluded region. The green area in the bottom is ruled out by Fermi-LAT Dwarf Spheroidals bounds. The black line reproduces $\Omega h^2 = 0.12$, whereas the regions below and beneath the lines set $\Omega h^2 > 0.12$ and $\Omega h^2 < 0.12$ respectively. }
\label{fig3}
\end{figure}

Assuming that the connection between the visible and the dark sector is given exclusively through the $Z^{\prime}$ portal with the coupling factors $a=b=0.5$ we have computed the abundance and the black line in Fig.\ref{fig4} reproduces $\Omega h^2 = 0.12$. The region beneath the black line provides $\Omega h^2 < 0.12$, whereas the one above the line $\Omega h^2 > 0.12$. Notice that setting $a=b=0.5$ means that we are taking the $Z^{\prime}-quarks$ suppressed in $50\%$ in comparison with the SM Z-quarks ones.

In Fig.\ref{fig4} we plot the parameter space $M_{Z^{\prime}} \times g_{\chi}$ allowed (white) and excluded (shaded and colorful) by the combined collider, direct and indirect detection bounds. Pink region is ruled out by the LHC $8$~TeV dijet data. Yellow is excluded by Tevatron ($1.96$~TeV) with the dijet data. The dashed blue region is excluded by LHC at $8$~TeV using the $jet+ E_T^{miss}$ data. The red (blue) dashed region is the XENON100 (LUX 2013) excluded region, whereas in green we show Fermi-LAT Dwarfs one. Notice that when we suppress the $Z^{\prime}$-quarks couplings in $50\%$ new regions of the parameters space show up. In particular, $1$~TeV mediators with $g_{\chi} \simeq 0.1$ are allowed by the data.
\begin{figure}[!t]
\centering
\includegraphics[scale=0.7]{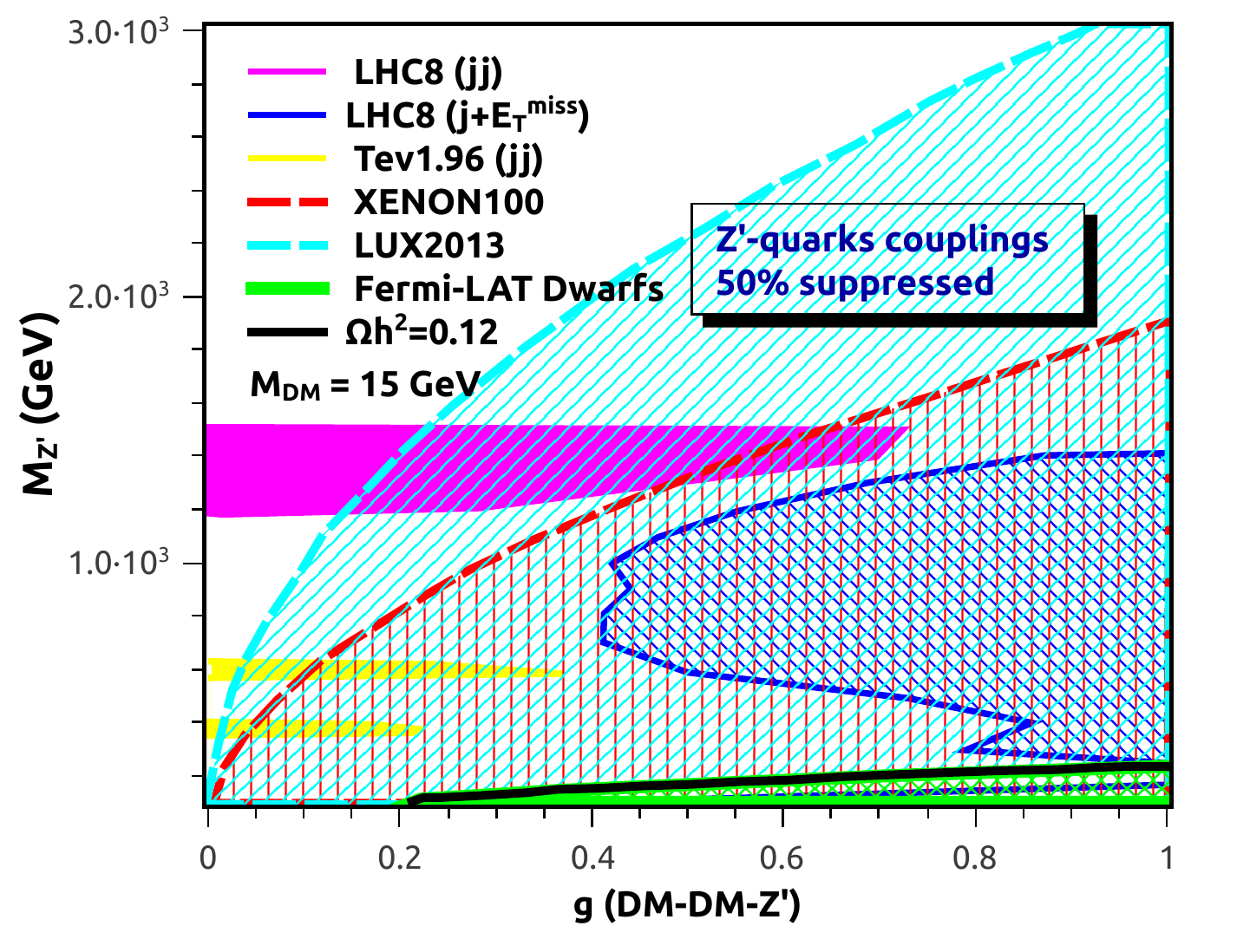}
\caption{Result for $\mchi=15$~GeV in the $M_{Z^{\prime}} \times$ DM-DM-$Z^{\prime}$ coupling plane, with $Z^{\prime}$-quarks couplings suppressed. The white regions are not ruled out by any constraint. The pink region is ruled out by the LHC $8$~TeV using the dijet data. Yellow is excluded by Tevatron ($1.96$ TeV) with the dijet data. The dashed blue region is excluded by LHC at $8$~TeV using the $jet+ E_T^{miss}$ analysis. The red dashed region is the XENON100 excluded region. The black line reproduces $\Omega h^2 = 0.12$, whereas the regions below and beneath the lines set $\Omega h^2 > 0.12$ and $\Omega h^2 < 0.12$ respectively.}
\label{fig4}
\end{figure}
\subsection{$130$~GeV WIMP in light of the Fermi line}

The $130$~GeV gamma-ray line observed in the Fermi-LAT data \cite{gammaline1,gammaline2} has appeared as a smoking gun signature for a $130$~GeV DM particle which has a fairly large annihilation cross section into $\gamma \gamma$. Different groups have found the same line feature in the Fermi data \cite{lineother}. The Fermi-LAT Collaboration, though, has claimed no evidence for such a line \cite{linefermi2}. The new PASS8 software plus additional accumulated data will hopefully shed some light on this interesting possibility  \cite{linefermi1}. Hereunder, we will be agnostic as to the particular particle mechanism responsible for the production of the line, and simply consider the particular value of 130 GeV as interesting for the dark matter particle mass. 

First, we show in Fig.\ref{figCS4} the spin independent cross section as a function of the $Z^{\prime}$ mass with $g_{\chi}$ varying randomly from zero to one. In the {\it left panel} we have assumed the coupling factors $a=b=1$, while in the in the {\it right panel} we have used a $50\%$ suppression in the $Z^{\prime}$-quarks couplings, i.e, with $a=b=0.5$. We can clearly see in the left and right panels of Fig.\ref{figCS4} the region between the blue and green point, which delimits the right abundance regime, is ruled out or on the verge of being excluded. Therefore a particle physics model which has a $Z^{\prime}$ only mediating the annihilation and scattering channels is rather disfavored by the current data.

In Fig.\ref{fig7} we have drawn a black line that sets $\Omega h^2 = 0.12$ for a $130$~GeV DM particle as a function of the DM-DM-$Z^{\prime}$ coupling ($g_{\chi}$) assuming that the connection between the visible and the dark sector is given through the $Z^{\prime}$ portal only with the coupling factors $a=b=1$, i.e, in the regime that the $Z^{\prime}$ boson has identical couplings to the SM Z. The area beneath the black curve provides $\Omega h^2 < 0.12$, whereas the one above the line $\Omega h^2 > 0.12$. Hence $Z^{\prime}$ mediated processes may provide the right abundance as long as $M_{Z^{\prime}} \lesssim 1$~TeV.

In Fig.\ref{fig7} we have shown as well the parameter space $M_{Z^{\prime}} \times g_{\chi}$ along with the collider, direct and indirect detection bounds.  The white regions are not ruled out by any constraint. Pink and Grey regions are ruled out by the LHC $15$~TeV  and $7$~TeV using the dijet data respectively. Yellow is excluded by Tevatron ($1.96$~TeV) with the dijet data. The dashed blue region is excluded by LHC at $15$~TeV using the $jet+ E_T^{miss}$ analysis. The red (blue) dashed region is the XENON100 (LUX 2013) excluded region, whereas the green one is excluded by Fermi-LAT Dwarfs data \cite{Fermidwarfs}. We can conclude from Fig.\ref{fig7} that the whole light mediator window is excluded and only the region with $M_{Z^{\prime}} \sim 1100$ for $g_{\chi} < 0.05$ circumvents the bounds. Apart from this one, only heavy mediators $M_{Z^{\prime}}> 1.7$~TeV, are allowed by the current data for $g_{\chi} < 0.3$. As we increase the mass larger couplings are allowed by data. For larger couplings the $Z^{\prime}$ mass is excluded up to $3$~TeV. 
\begin{figure}[!t]
\centering
\includegraphics[scale=0.55]{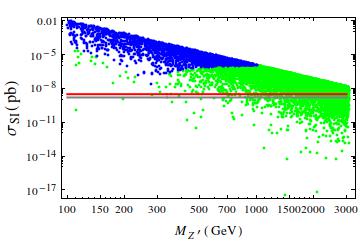}
\includegraphics[scale=0.55]{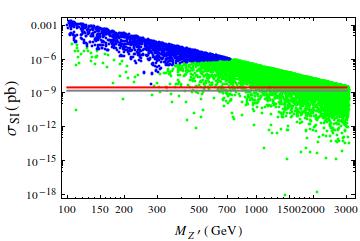}
\caption{Spin independent cross section as a function of the $Z^{\prime}$ for $\mchi=130$~GeV with $a=b=1 (left)$ and $a=b=0.5 (right)$.  Green points provide $\Omega h^2 > 0.12$ (overabundant), whereas the blue ones $\Omega h^2 < 0.11$ (underabundant). The horizontal lines are the XENON100 (top) and LUX (bottom) limits.}
\label{figCS4}
\end{figure}
\begin{figure}[!t]
\centering
\includegraphics[scale=0.7]{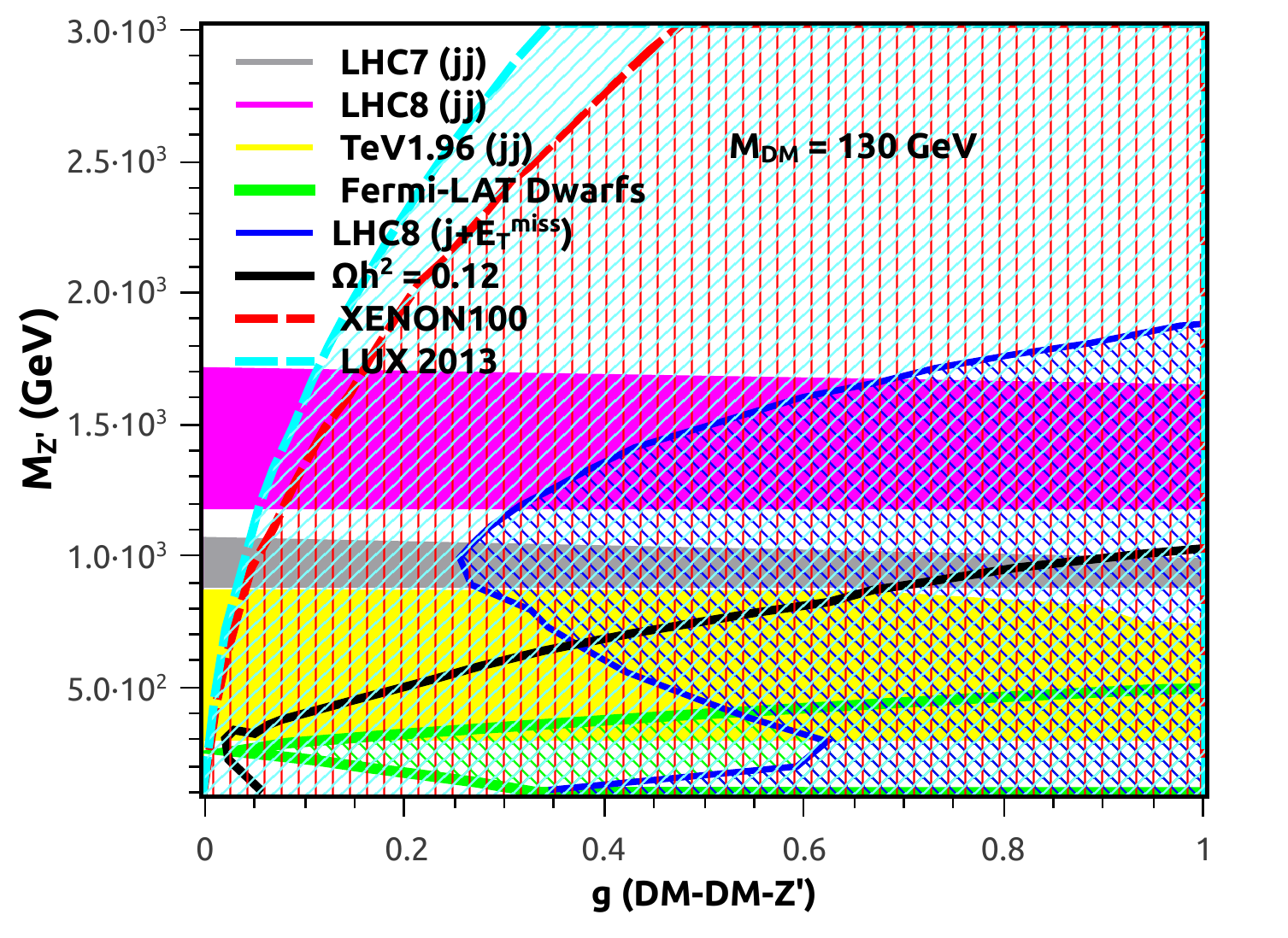}
\caption{Result for $\mchi=130$~GeV in the $M_{Z^{\prime}} \times$ DM-DM-$Z^{\prime}$ coupling plane. The white regions are not ruled out by any constraint. The gaps obey all constraints. Pink and Grey regions are ruled out by the LHC $8$~TeV  and $7$~TeV using the dijet data. Yellow is excluded by Tevatron ($1.96$ TeV) with the dijet data. The dashed blue region is excluded by LHC at $8$~TeV using the $jet+ E_T^{miss}$ analysis. The red dashed region is the XENON100 excluded region. The black line reproduces $\Omega h^2 = 0.12$, whereas the areas below and beneath the black curve set $\Omega h^2 > 0.12$ and $\Omega h^2 < 0.12$ respectively.}
\label{fig7}
\end{figure}
Now in Fig.\ref{fig8} we show the impact of suppressing in $50\%$ the $Z^{\prime}$-quarks couplings with  $\mchi=130$~GeV. It is clear from Fig.\ref{fig8} that a small window for $M_{Z^{\prime}} \sim 500$~GeV with suppressed couplings opens up and a somewhat large one for
$700\ \mbox{GeV} < M_{Z^{\prime}} < 1.2$~TeV with $g_{\chi} < 0.1$ becomes allowed by the joined data. Besides these only heavy mediator are consistent with the current data. 

\begin{figure}[!t]
\centering
\includegraphics[scale=0.7]{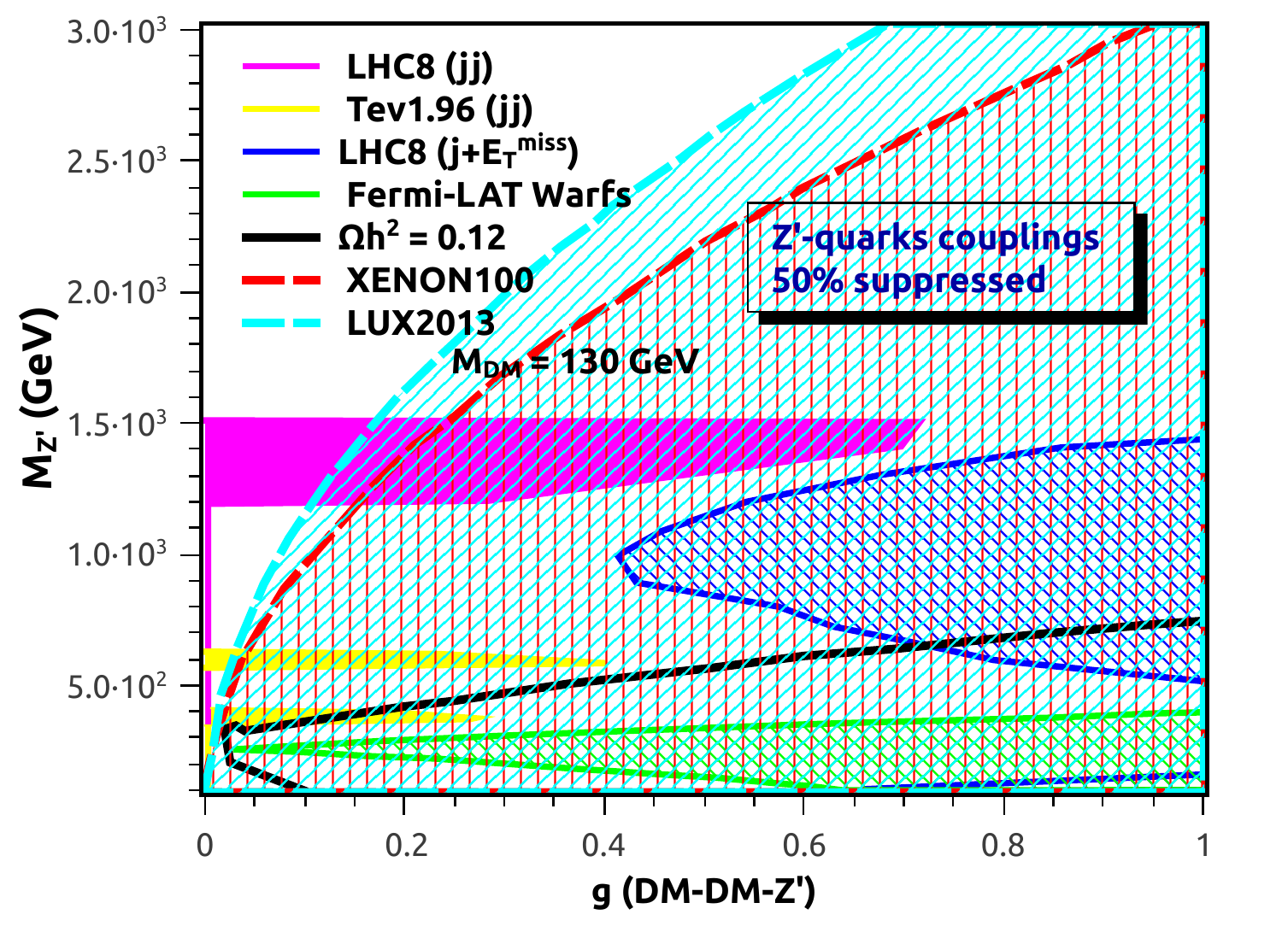}
\caption{Result for $\mchi=130$~GeV in the $M_{Z^{\prime}} \times$ DM-DM-$Z^{\prime}$ coupling plane, with $Z^{\prime}$-quarks couplings suppressed. The white regions are not ruled out by any constraint. The pink region is ruled out by the LHC $8$~TeV using the dijet data. Yellow is excluded by Tevatron ($1.96$ TeV) with the dijet data. The dashed blue region is excluded by LHC at $8$~TeV using the $jet+ E_T^{miss}$ analysis. The red dashed region is the XENON100 excluded region. The black line reproduces $\Omega h^2 = 0.12$, whereas the regions below and beneath the black curve set $\Omega h^2 > 0.12$ and $\Omega h^2 < 0.12$ respectively.}
\label{fig8}
\end{figure}

\subsection{Results for $\mchi=50,500,1000$~GeV WIMPs}

In this final section we consider three additional values for the WIMP mass: 50, 500 and 1000 GeV, and study how our results extrapolate to larger masses. The structure of all plots is the same as before: for each value of the mass we show the spin independent cross section as a function of $m_{\zp}$ for $a=b=1$ (left) and for $a=b=0.5$ (right). We then show the over- and under-abundant thermal relic density regions and the bounds from direct, indirect and collider searches both for suppressed and unsuppressed quark couplings.

While nothing qualitatively new emerges for the $\mchi=50$ GeV case, in the $\mchi=500$ and 1000 GeV case we notice the new feature associated with the presence of a resonance at $\mchi=m_{\zp}/2$. While portions of that resonance are highly constrained by collider searches and by indirect searches (the resonance enhances the pair-annihilation cross section via an s-channel $\zp$ exchange), we find interesting portions of parameter space that are viable and that feature the correct thermal relic density.

The feature we highlighted for the 500 and 1000 GeV mass persists for even larger masses, where the portion of parameter space compatible with the thermal relic density requirement and with collider searches continues to be present. The only constraint is given by gamma-ray searches, which benefit from the large annihilation cross sections associated with the resonant annihilation mode.

\begin{figure}[!t]
\centering
\includegraphics[scale=0.55]{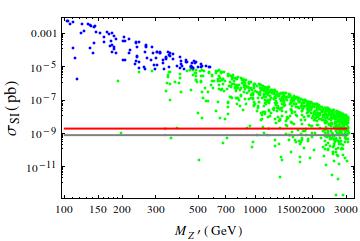}
\includegraphics[scale=0.55]{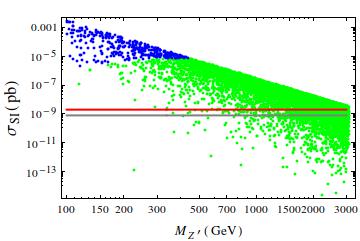}
\caption{Spin independent cross section as a function of the $Z^{\prime}$ for $\mchi=50$~GeV with $a=b=1 (left)$ and $a=b=0.5 (right)$.  Green points provide $\Omega h^2 > 0.12$ (overabundant), whereas the blue ones $\Omega h^2 < 0.11$ (underabundant). The horizontal lines are the XENON100 (top) and LUX (bottom) limits. }
\label{figCS3}
\end{figure}

\begin{figure}[!t]
\centering
\includegraphics[scale=0.7]{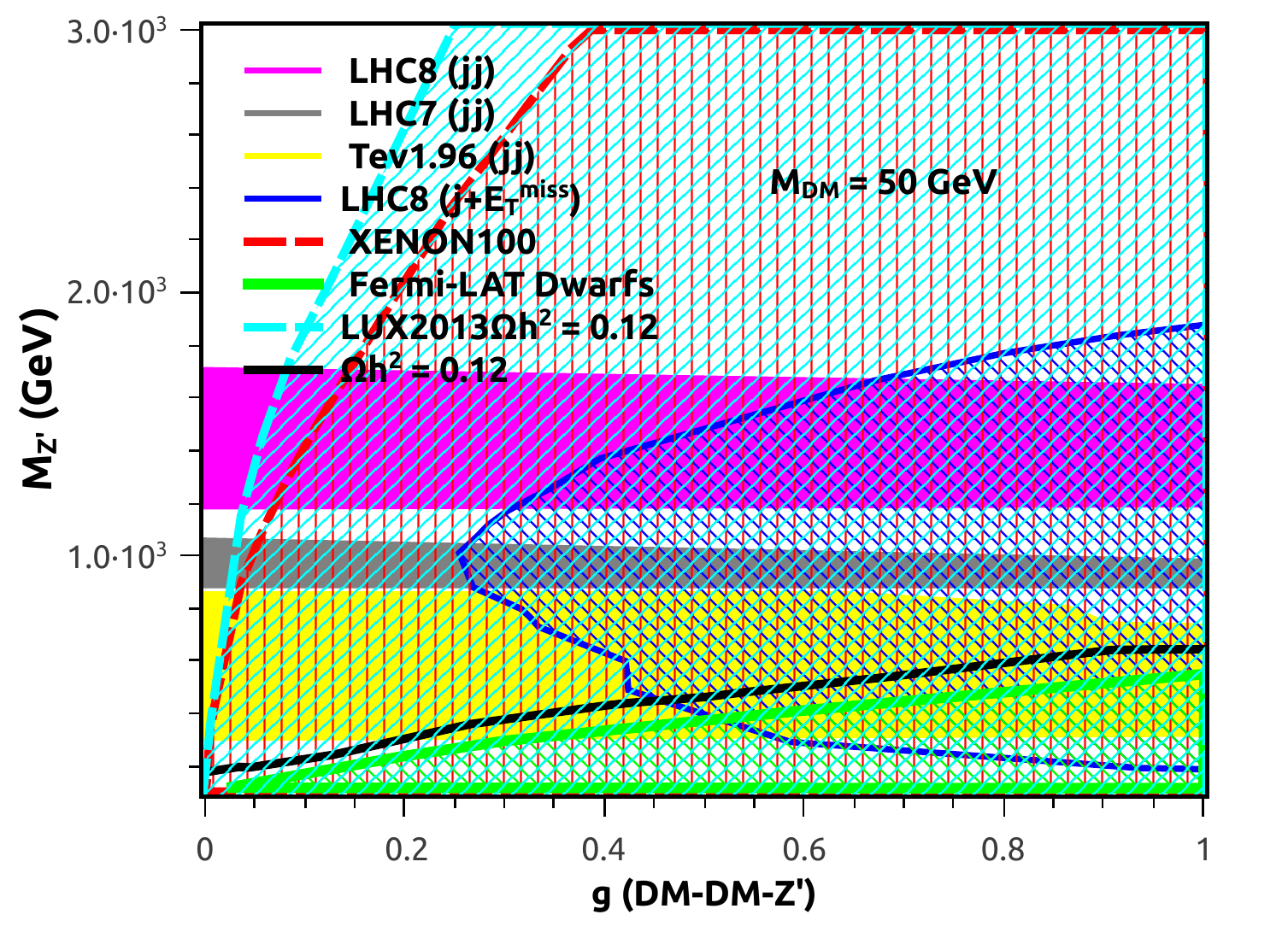}
\caption{Result for $\mchi=50$~GeV in the $M_{Z^{\prime}} \times$ DM-DM-$Z^{\prime}$ coupling plane. The white regions are not ruled out by any constraint. Pink and Grey regions are ruled out by the LHC $8$~TeV  and $7$~TeV using the dijet data. Yellow is excluded by Tevatron ($1.96$ TeV) with the dijet data. The dashed blue region is excluded by LHC at $8$~TeV using the $jet+ E_T^{miss}$ analysis. The red dashed region is the XENON100 excluded region. The black line reproduces $\Omega h^2 = 0.12$, whereas the regions below and beneath the lines set $\Omega h^2 > 0.12$ and $\Omega h^2 < 0.12$ respectively.}
\label{fig5}
\end{figure}

\begin{figure}[!t]
\centering
\includegraphics[scale=0.7]{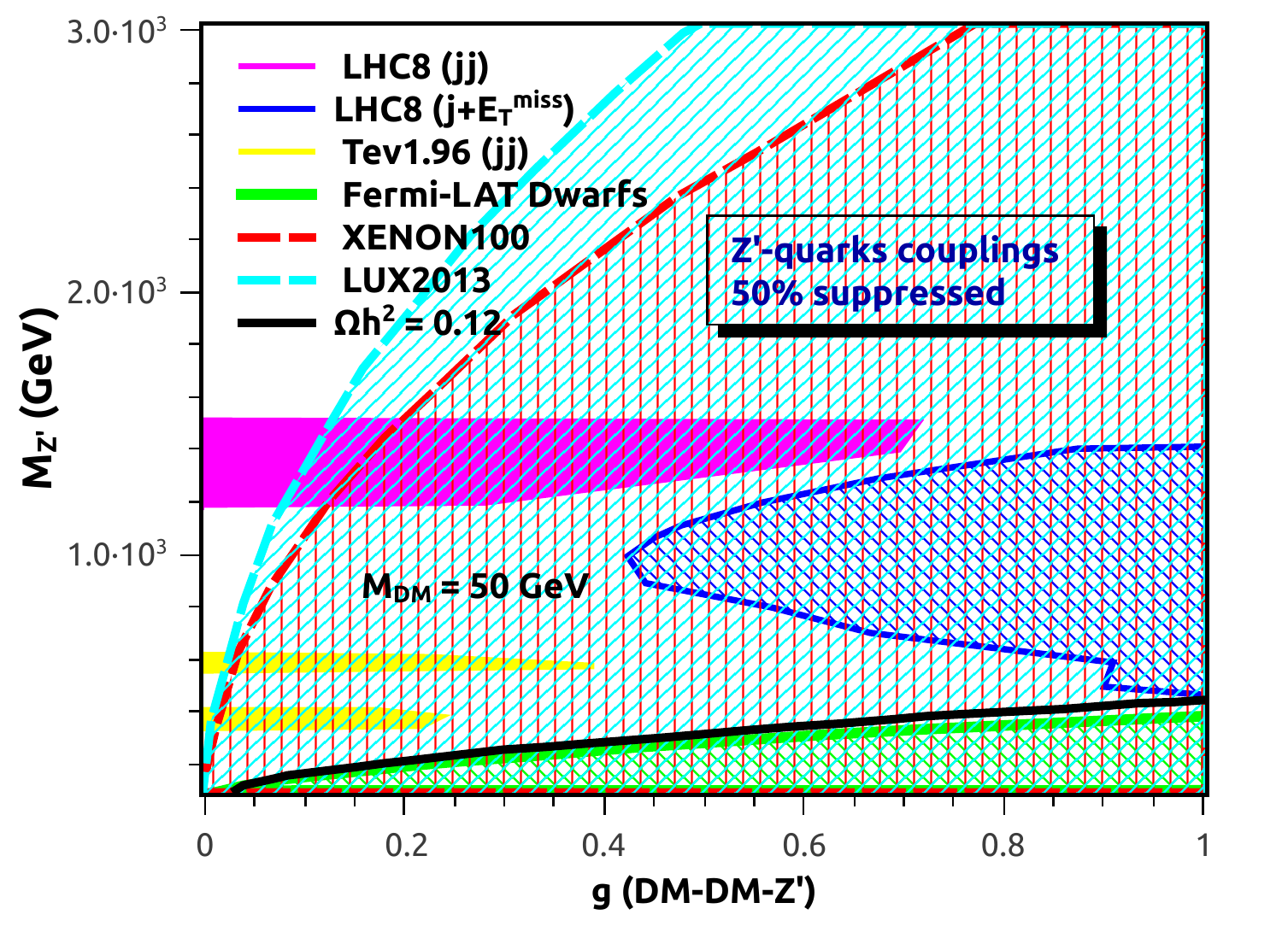}
\caption{Result for $\mchi=50$~GeV in the $M_{Z^{\prime}} \times$ DM-DM-$Z^{\prime}$ coupling plane, with $Z^{\prime}$-quarks couplings suppressed. The gaps evade all bounds. The pink region is ruled out by the LHC $8$~TeV using the dijet data. Yellow is excluded by Tevatron ($1.96$ TeV) with the dijet data. The dashed blue region is excluded by LHC at $8$~TeV using the $jet+ E_T^{miss}$ analysis. The red dashed region is the XENON100 excluded region. The black line reproduces $\Omega h^2 = 0.12$, whereas the regions below and beneath the lines set $\Omega h^2 > 0.12$ and $\Omega h^2 < 0.12$ respectively.}
\label{fig6}
\end{figure}


\begin{figure}[!t]
\centering
\includegraphics[scale=0.55]{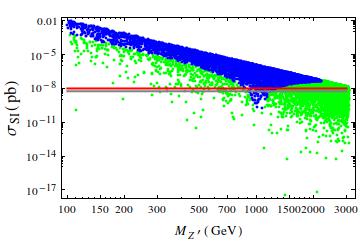}
\includegraphics[scale=0.55]{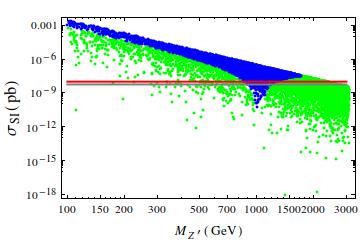}
\caption{Spin independent cross section as a function of the $Z^{\prime}$ for $\mchi=500$~GeV with $a=b=1 (left)$ and $a=b=0.5 (right)$.  Green points provide $\Omega h^2 > 0.12$ (overabundant), whereas the blue ones $\Omega h^2 < 0.11$ (underabundant). The horizontal lines are the XENON100 (top) and LUX (bottom) limits. }
\label{figCS5}
\end{figure}

\begin{figure}[!t]
\centering
\includegraphics[scale=0.7]{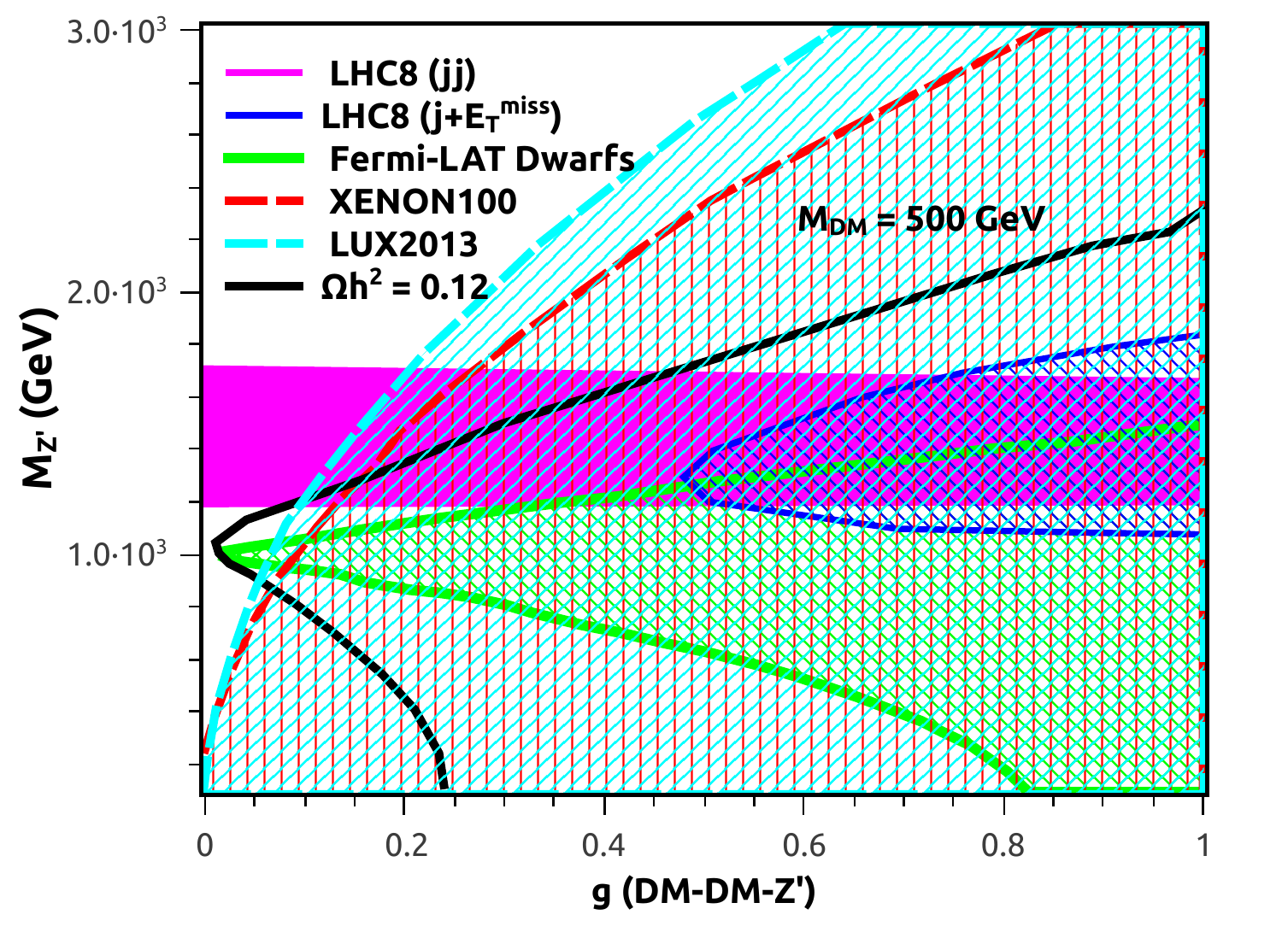}
\caption{Result for $\mchi=500$~GeV in the $M_{Z^{\prime}} \times$ DM-DM-$Z^{\prime}$ coupling plane. The gaps evade all bounds. 
The pink region is ruled out using the dijet data from the LHC 8 TeV.
The dashed blue region is excluded by LHC at $8$~TeV using the $jet+ E_T^{miss}$ analysis. The red dashed region is the XENON100 excluded region. The black line reproduces $\Omega h^2 = 0.12$, whereas the regions below and beneath the lines set $\Omega h^2 > 0.12$ and $\Omega h^2 < 0.12$ respectively.}
\label{fig8_1}
\end{figure}

\begin{figure}[!t]
\centering
\includegraphics[scale=0.7]{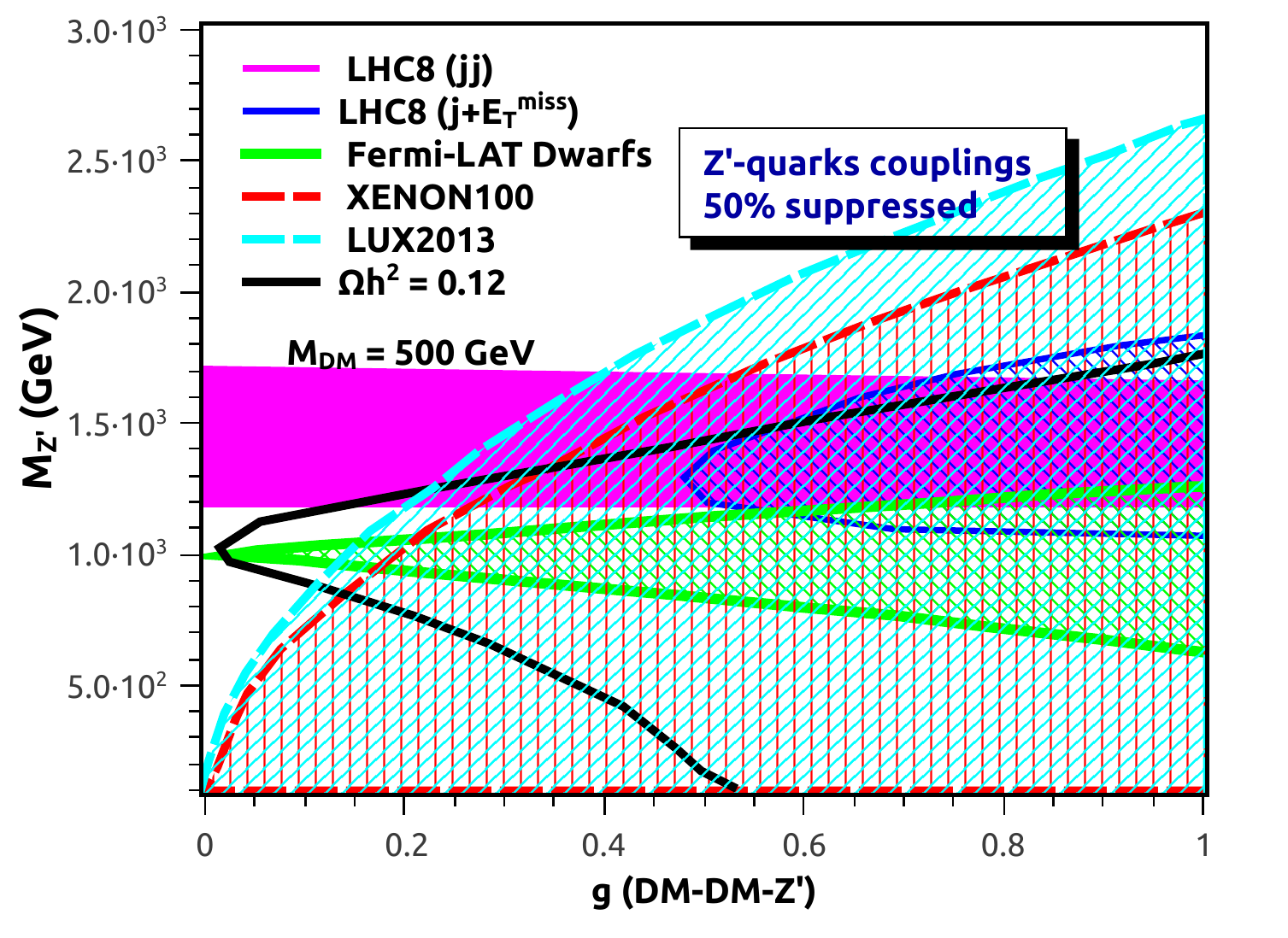}
\caption{Result for $\mchi=500$~GeV in the $M_{Z^{\prime}} \times$ DM-DM-$Z^{\prime}$ coupling plane, with $Z^{\prime}$-quarks couplings suppressed. The gaps evade all bounds. 
The pink region is ruled out using the dijet data from the LHC 8 TeV. 
The dashed blue region is excluded by LHC at $8$~TeV using the $jet+ E_T^{miss}$ analysis. The red dashed region is the XENON100 excluded region. The black line reproduces $\Omega h^2 = 0.12$, whereas the regions below and beneath the lines set $\Omega h^2 > 0.12$ and $\Omega h^2 < 0.12$ respectively.}
\label{fig8_2}
\end{figure}


\begin{figure}[!t]
\centering
\includegraphics[scale=0.55]{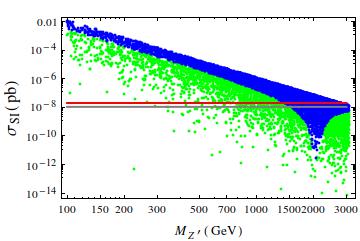}
\includegraphics[scale=0.55]{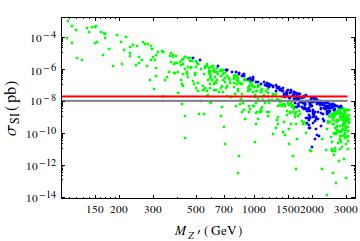}
\caption{Spin independent cross section as a function of the $Z^{\prime}$ for $\mchi=1000$~GeV with $a=b=1 (left)$ and $a=b=0.5 (right)$.  Green points provide $\Omega h^2 > 0.12$ (overabundant), whereas the blue ones $\Omega h^2 < 0.11$ (underabundant). The horizontal lines are the XENON100 (top) and LUX (bottom) limits. }
\label{figCS6}
\end{figure}

\begin{figure}[!t]
\centering
\includegraphics[scale=0.6]{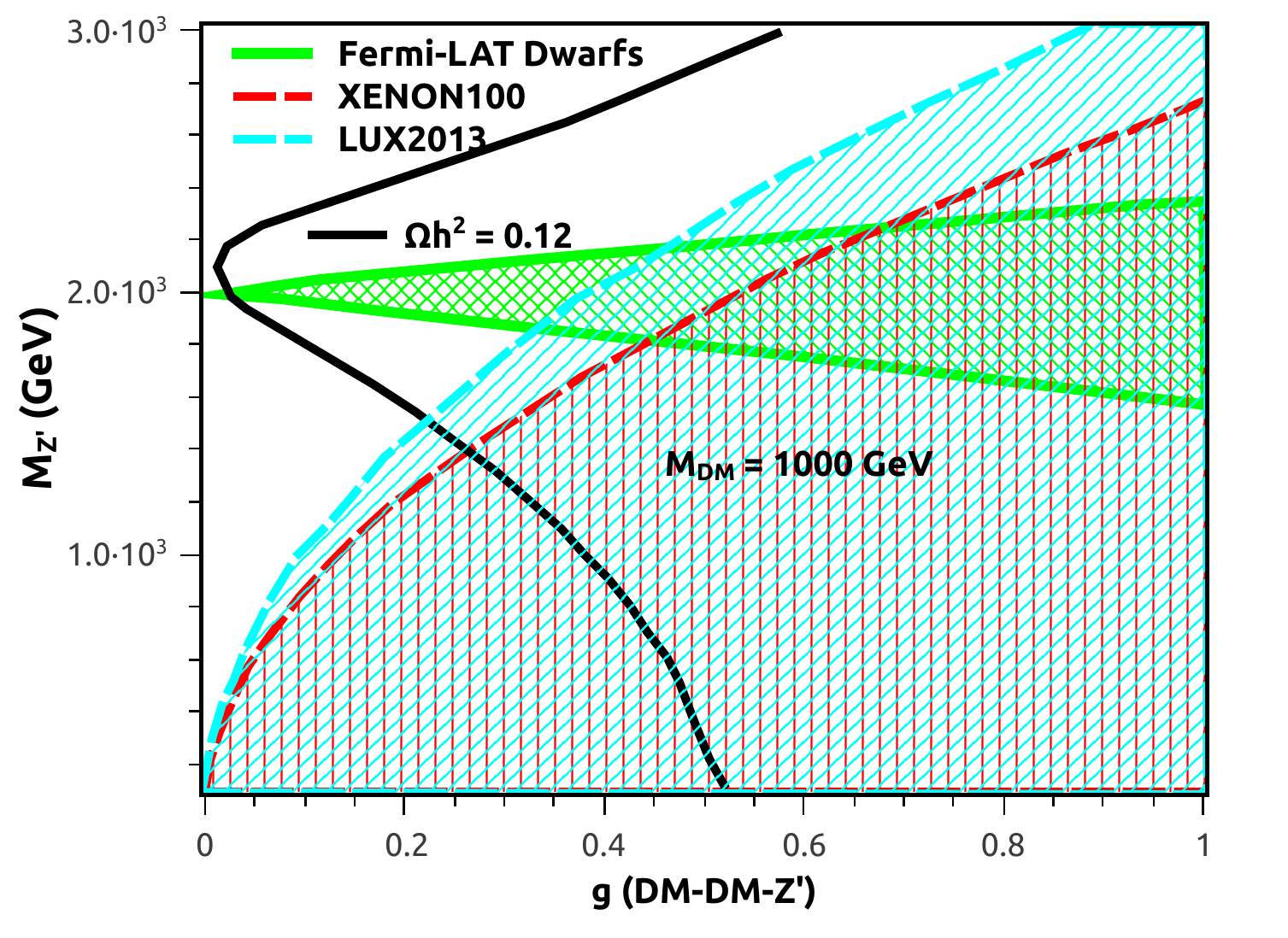}
\caption{Result for $\mchi=1000$~GeV in the $M_{Z^{\prime}} \times$ DM-DM-$Z^{\prime}$ coupling plane.  The gaps evade all bounds. The pink region is ruled out by the LHC $8$~TeV.  
The red dashed region is the XENON100 excluded region. The black line reproduces $\Omega h^2 = 0.12$, whereas the regions below and beneath the lines set $\Omega h^2 > 0.12$ and $\Omega h^2 < 0.12$ respectively.}
\label{fig9}
\end{figure}

\begin{figure}[!t]
\centering
\includegraphics[scale=0.6]{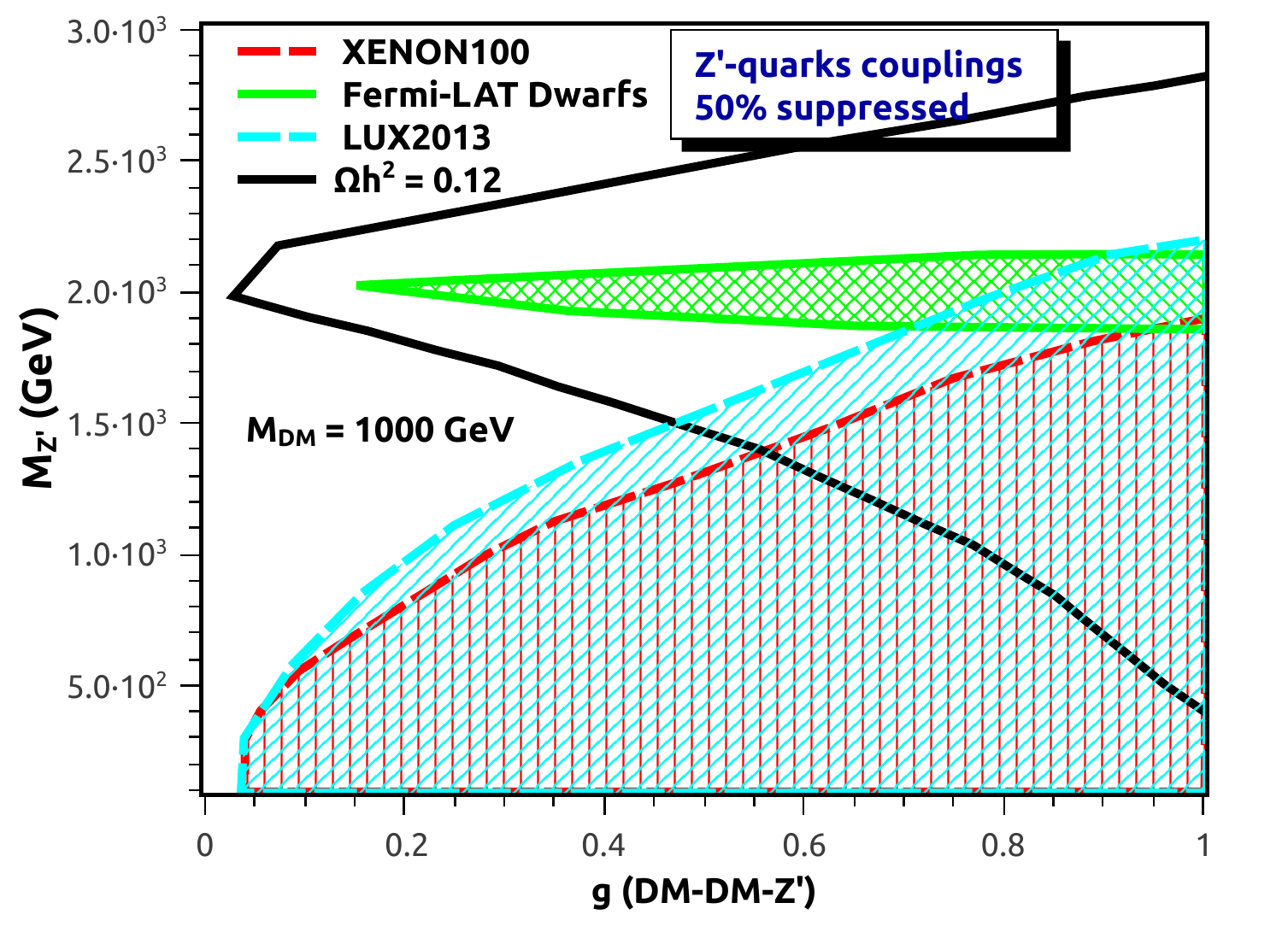}
\caption{Result for $\mchi=1000$~GeV in the $M_{Z^{\prime}} \times$ DM-DM-$Z^{\prime}$ coupling plane, with $Z^{\prime}$-quarks couplings suppressed. The white regions are not ruled out by any constraint. 
The red dashed region is the XENON100 excluded region. The black line reproduces $\Omega h^2 = 0.12$, whereas the regions below and beneath the lines set $\Omega h^2 > 0.12$ and $\Omega h^2 < 0.12$ respectively.}
\label{fig10}
\end{figure}

\section{Conclusions}
In this study we have investigated a particular incarnation of the $\zp$ dark portal, where the additional gauge boson does not interact with leptons. We carried out a detailed study of direct and indirect searches for the resulting dark matter candidate, as well as an extensive study of the collider phenomenology using Tevatron and LHC results.

We found a high degree of complementarity at several different levels: between direct and indirect dark matter searches, between dark matter searches and collider studies, and between Tevatron and LHC searches. We focused our detailed analyses on specific values of the dark matter particle mass, motivated by tentative signals in both direct and indirect detection. Inspecting the thermal relic density we found that only for large masses can we obtain regions that are not ruled out by dark matter or collider searches and that possess the correct universal abundance. In those regions, the mass of the dark matter is about half the mass of the $\zp$ and resonant annihilation produces large cross sections contributing to suppressing the relic density, with other parameters compatible with direct and collider searches. 

\section*{Acknowledgments}
This work is partly supported by the Department of Energy under contract DE-FG02-04ER41286 (SP), and by the Conselho Nacional de Desenvolvimento Cient\'ifico e Tecnol\'ogico (CNPq) (FSQ). The authors thank Patrick Draper, William Shepherd and Chris Kelso for valuable comments.

\end{document}